\documentclass[twocolumn]{aastex701}

\usepackage{amsmath}
\usepackage{array}
\usepackage{multirow}
\newcolumntype{L}[1]{>{\raggedright\arraybackslash}p{#1}}

\usepackage[mathscr]{euscript}
 \let\mathscr\relax
\usepackage[scr]{rsfso}

\begin{document}

\title{Do Rocky Planets Around M Stars Have Atmospheres? \\A Statistical Approach to the Cosmic Shoreline}

\correspondingauthor{Jegug Ih}
\email{jih@stsci.edu}

\author[0000-0003-2775-653X]{Jegug Ih}
\affiliation{Space Telescope Science Institute, 3700 San Martin Drive, Baltimore, MD 21218, USA}
\email{jih@stsci.edu}

\author[0000-0002-1337-9051]{Eliza M.-R. Kempton}
\affiliation{Department of Astronomy \& Astrophysics, University of Chicago, Chicago, IL 60637, USA}
\affiliation{Department of Astronomy, University of Maryland, College Park, MD 20742, USA}
\email{}

\author[0000-0001-8274-6639]{Hannah Diamond-Lowe}
\affiliation{Space Telescope Science Institute, 3700 San Martin Drive, Baltimore, MD 21218, USA}
\email{}

\author[0000-0001-6878-4866]{Joshua Krissansen-Totton}
\affiliation{Department of Earth and Space Sciences, University of Washington, Seattle, WA 98195, USA}
\email{}

\author[0000-0003-4241-7413]{Megan Weiner Mansfield}
\affiliation{School of Earth and Space Exploration, Arizona State University, Tempe, AZ 85287, USA}
\affiliation{Department of Astronomy, University of Maryland, College Park, MD 20742, USA}
\email{}

\author[0000-0002-6215-5425]{Qiao Xue}
\affiliation{Department of Astronomy \& Astrophysics, University of Chicago, Chicago, IL 60637, USA}
\email{}

\author[0000-0002-0413-3308]{Nicholas Wogan}
\affiliation{NASA Ames Research Center, Mountain View, CA 94035, USA}
\email{}

\author[0000-0001-8236-5553]{Matthew C. Nixon}
\affiliation{Department of Astronomy, University of Maryland, College Park, MD 20742, USA}
\email{}

\author[0000-0001-5084-4269]{Benjamin J. Hord}
\affiliation{NASA Goddard Space Flight Center, Greenbelt, MD 20771, USA}
\email{}



\begin{abstract}

Answering the question ``do rocky exoplanets around M stars have atmospheres?'' is a key science goal of the JWST mission, with 500 hours of Director’s Discretionary Time (DDT) awarded to address it. Theoretically, the so-called ``Cosmic Shoreline'' may not hold around M stars due to their harsher XUV environment, possibly resulting in most rocky planets lacking significant atmospheres---a hypothesis that remains to be statistically tested through judicious target selection.  We identify target selection as a combinatorial optimization problem (``knapsack problem'').  We develop a statistical framework to test population-level hypotheses from observations and combine a formation and evolution model, 1D-RCE atmosphere model, and genetic algorithm to simulate populations and find the optimized set of observations.  We find that, firstly, if all rocky planets around M stars are indeed bare rocks, JWST can efficiently place an upper bound on the atmosphere occurrence rates to less than 1 in 8, even without optimized target selection, but further improvements are cost-prohibitive.  Secondly, if the Cosmic Shoreline hypothesis (XUV or bolometric) holds true for M stars, strong evidence ($\Delta$BIC$>5$) can be found within $\sim500$ observing hours using the optimal strategy of a ``wide and shallow'' approach.  Our statistical framework can be directly applied to upcoming observations to robustly identify the Cosmic Shoreline and to optimize target selection for determining other trends in exoplanet atmosphere observations, including those from future missions.  

\end{abstract}



\section{Introduction} \label{sec:intro}










\subsection{The M-Star Cosmic Shoreline}

One key scientific objective of the James Webb Space Telescope (JWST) is to determine whether and under what conditions terrestrial exoplanets can form and retain their atmospheres.  Answering this question takes the first necessary step towards understanding habitability in systems outside the solar system and will be a lasting legacy of JWST that builds the groundwork for future missions that aim to directly detect biosignatures \citep{decadal2020}.  For this question, observing transiting rocky exoplanets around M stars is particularly useful, as these stars are the most abundant in the solar neighborhood and their smaller stellar radii and cooler stellar temperature yield favorable signal sizes. 

Indeed, the main workhorse for identifying whether a rocky planet (i.e., smaller than $\sim$1.7 $R_{\mathrm{\oplus}}$) \citep{rogers15} has an atmosphere in the first few years of JWST has been to observe targets orbiting M stars in secondary eclipse, \textit{i.e.} observing the star as the planet moves in and out of view behind the star, using the Mid-Infrared Instrument (MIRI).  These observations allow for measuring the dayside thermal emission of the planet and thereby measure the dayside temperature and energy budget.  As a thick atmosphere is expected to cool the dayside by redistributing heat from the incident stellar flux away to the nightside, the presence and extent of an atmosphere can be inferred via comparing the observed temperature to the expected value for a bare low-albedo rock, if the system parameters (orbital distance and stellar effective temperature) are known \citep{mansfield19, koll19}.  Additionally, if observations are made spectroscopically or in multiple bands, the additional information can in theory be used to simultaneously constrain the composition and the surface pressure of the atmosphere \citep{deming09, whittaker22}.  Secondary eclipse observations are advantageous over transmission observations in that they are largely not affected by stellar contamination or degeneracy with clouds, which obfuscates interpretation and may inhibit stacking multiple transits \citep{rackham18, lustig19}.


This secondary eclipse technique has been successfully used to rule out thick ($\geq 1$ bar), CO$_2$-rich atmospheres on planets using \textit{Spitzer} and JWST, using both MIRI photometry and MIRI LRS spectroscopy.  Examples include LHS 3844 b \citep{kreidberg19}, TRAPPIST-1 b \citep{greene23, ducrot23}, GJ 1132 b \citep{xue24}, Gl 486 b \citep{mansfield24}, TOI-1468 b \citep{meier2025_toi1468b}, LHS 1140 c \citep{fortune25_lhs1140c}.  So far, no definitive detection of an atmosphere has been made, and all eclipse observations have been consistent with a bare blackbody rock to varying degrees.  On the other hand, transmission observations have indicated tentative detections of sulfur-dominated atmospheres on the L 98-59 system on planets b and c, for instance; however, the detections of the atmospheres (based on atmospheric absorption features) remain of low significance \citep{gressier24, bello25}.  In light of these observations, we can reasonably ask whether any M-dwarf rocky planets have atmospheres at all, and if so, what governs their presence or absence.

Studying the targets at the \textit{population level} can allow for synthesizing these observations into meaningful statements about their nature, even from individually weak or null results \citep[e.g.,][]{bean17_statistical}.  Demonstrating the population-level approach, \citet{coy24} found a tentative trend in the observed brightness temperature of rocky exoplanets (normalized to the theoretical maximum dayside temperature) as a function of  irradiation temperature, which could either be explained by varying surface albedos with irradiation temperature due to, e.g., space weathering, or the onset of very tenuous atmospheres at lower temperatures.  Given the substantial number of potential targets available to JWST (Figure \ref{fig:cosho}), capitalizing on the population-level approach is a promising avenue to understanding the prevalence of atmospheres, as demonstrated in, e.g., the JWST Cycle 2 Hot Rocks Survey program (PI: Diamond-Lowe, GO 3730).

From a formation and evolution point of view, it remains unclear whether rocky planets around M stars are likely to retain their atmospheres, as a complex tapestry of competing processes and factors exists.   Several pathways could lead to the complete loss of atmospheres, including thermal escape \citep{tian09}, loss of high mean molecular volatiles during hydrodynamic escape of primary atmospheres \citep{kite20, krissansentotton24_erosion}, non-thermal escape such as impact-induced stripping \citep{wyatt19_impact} or stellar wind interactions \citep{dong18}, and volatile-poor initial formation \citep{lissauer07_planets}.  Conversely, several mechanisms could contribute to atmospheric \mbox{(re-)generation} or retention; processes for the former include outgassing from a magma ocean, volcanic replenishment, or late-stage delivery of volatiles; while the latter include efficient atomic line cooling in the upper atmosphere that inhibits thermal escape \citep{nakayama22, chatterjee24}.

In the Solar System, the so-called ``Cosmic Shoreline'' is the quasi-empirical trend that separates airless bodies from those with atmospheres in the escape speed-instellation plane.  Such a correlation is consistent with a picture in which energy-limited escape (or other escape mechanisms) determine which planets retain atmospheres \citep{zahnle17}.  It remains to be tested whether the mosaic of processes governing rocky planet atmospheres reduces to a similarly simple correlation for planets orbiting M stars; in this sense, the population-level question of rocky planet atmospheres around M stars can be framed as determining whether the Cosmic Shoreline concept extends to M-dwarf planets and constraining its position and shape.

If atmospheric escape is indeed the primary driver that carves out the Cosmic Shoreline for M stars, the radiation environment around M stars is different in two critical ways compared to more massive stars.  Firstly, M stars emit a larger fraction of their bolometric flux in the X-ray and extreme ultraviolet (XUV) range than Sun-like stars, which drives thermal atmospheric escape \citep{shields16, zhu25}.  Secondly, M stars have longer pre-main-sequence lifetimes and spin-down history, implying that the planets have been irradiated for much longer.  As such, even planets at similar current (bolometric) instellations to Earth or Venus may have experienced more escape if they orbit M stars  \citep{luger15, vanlooveren24, vanlooveren25}.  This is especially true for \textit{late} M stars (spectral subtype later than 3.5) that are fully convective and have a much longer spin-down history \citep{wright11, charbonneau23, pass25}.  However, if the Cosmic Shoreline requires a modification for M stars in some form, an additional dependence on the host stellar temperature as the third dimension is a strong possibility.

\subsection{Precisely defining the science goal}

Careful target selection is critical for a successful survey in search of a population-level trend \citep{bean17_statistical, batalha23_importance}.  \citet{batalha23_importance}, in particular, found that simply choosing the best targets by signal-to-noise (S/N) metric, \textit{i.e.} choosing the targets that would produce the best individual results, may not be the best set of targets that constrain a given property at the population-level.  As such, the specific science goal of interest must inform target selection.  This is especially true as how much telescope time is invested is not determined by the (a priori unknown) true nature of the target, but instead by what we predict for a set of targets based on what the specific science question is.  


In this work, we address the target selection problem for the science question: \textit{do rocky planets around M stars have atmospheres?} To answer this question quantitatively, we must first clarify what we actually mean by this question.  We propose the following three formulations:

\begin{itemize}
    \item Does at least one M-star rocky planet have an atmosphere?
    \item If all M-star rocky planets are indeed bare rocks, can we conclude this from observations?
    \item Is there a trend, similar to the Solar System Cosmic Shoreline, in which certain M-star rocky planets are more likely to have atmospheres?
\end{itemize}

While these three questions are not directed at entirely orthogonal science goals, they lead to three different approaches with different target selection priorities.   The first requires focusing observations to characterize promising systems individually.  This leads to a survey that is ``deep and narrow''.  For a definitive detection of an atmosphere without degenerate explanations, one would need to first detect a shallow eclipse and possibly follow up in another wavelength band or with a phase curve to robustly distinguish between an atmosphere redistributing heat and a false positive due to a bright surface \citep{hammond2024}.  Without any prior observation and with only the Cosmic Shoreline hypothesis to guide which targets are promising candidates, this becomes observationally expensive. For instance, if the 4 targets with the best Priority Metric (defined in \S \ref{subsec:targets}) each had 0.1 bar CO$_2$-dominated atmospheres,  as we will show in \S \ref{sec:discussion}, confidently making such a detection requires many thousands of hours of observing time, and runs the risk of producing null results in the end.  


The second question tries to show the negative of the first.  As one cannot \textit{prove} an inductive statement, this question would necessarily be answered statistically and thus becomes about placing an upper bound on an occurrence rate. As ruling out thick atmospheres is possible at a lower observational cost than confirming them, designing a survey around the a priori assumption that M-star rocky planets are likely to be bare rocks results in a ``wide and shallow'' survey, leveraging the statistical opportunity presented by the sample.  While one might surmise that a negative outcome to a survey designed to answer the first question, such as if all 4 of the best Priority Metric targets were revealed to be bare rocks, would allow for extrapolating that the rest of the sample are also bare rocks, we argue that this presupposes the M-star Cosmic Shoreline, which needs to be observationally tested.

The third question requires testing population-level hypotheses and evaluating their support (or rejection), which may be best constrained at the population level even if each target is weakly characterized individually \citep{coy24}.  Careful target selection is most important in answering this question, as not only do targets come with different costs in observing time but also contribute different values to the testing of the hypothesis based on their location in the parameter space of interest.  We explore the implications of framing the target selection in this manner in the following section.


\subsection{Target selection as an optimization problem}

In the current study, we frame the target selection as solving a non-linear knapsack problem, adopting the computer science jargon.  A \textit{knapsack problem} is a combinatorial optimization problem in which one must select a set of items, each with different weight and value, in order to maximize the total value whilst obeying some total weight limit \citep{martello1990knapsack}.  In this framing, each potential observation is an item that incurs a weight in observing time; the total weight is the total observing time to be kept under some limit; the total value is determined by how well the observations allow for statistically distinguishing different hypotheses.

In particular, target selection is of the \textit{non-linear} variety of the knapsack problem \citep{kellerer2013knapsack}, one that is non-linear in at least two ways.  Firstly, the total value does not equate to a mere sum of the component values but is instead determined by the synergy among the chosen items.  Secondly, there are generally diminishing returns to choosing the same object multiple times, as signal-to-noise scales less than linearly with the number of eclipses.  Because of this, the value of an individual item---a single eclipse observation of a target---is not exactly defined in our framing, as it strongly depends on what items have already been selected.

This non-linearity leads to interesting behaviors.  One key feature is that there is no \textit{optimal substructure} to the problem, meaning that the optimal solution to a large problem may not always contain the optimal sub-solution to a sub-problem.  For example, the best set of targets for a survey under 500 hours may not necessarily include the best set of targets for a survey under 100 hours.

Due to the lack of optimal substructure, \textit{greedy approaches}, in which one breaks down the full problem to sub-problems and grows a sub-solution recursively into the full solution, may not work.  In the previous example, the best set of targets for a 500-hour survey may not equate to the union of the best sets for 5 consecutive 100-hour surveys.  Because of this, should one try to answer the population-level question with a series of consecutive smaller self-contained surveys, it may potentially take longer than what one could achieve with a guaranteed amount of large time from the start.  Especially, in practice, one would design each mini-survey to provide self-contained results and thereby further deviate from focusing on the global population-level question.  This is doubly so for small planets with low S/N, where the number of stacked eclipses required for significant results are usually large.   Given these considerations, the 500 hours of Rocky Worlds DDT provides a unique opportunity to select the set of targets that \textit{efficiently} answers the population-level question.

\subsection{Purpose and structure of this work}

In this paper, we perform two experiments to address two key questions.  Firstly, should all M-star rocky planets indeed be bare rocks, we aim to quantify what constraints on the occurrence rates of atmospheres are achievable with a given amount of observing time.  Secondly, should the M-star Cosmic Shoreline exist in some form, we aim to establish how well the hypothesis can be statistically distinguished and identify the optimal target selection and observation strategies.

We note that our goal here is not to produce the definitive list of best targets specifically for the Rocky Worlds DDT; such a list should necessarily reflect the subjective answers to variegated considerations not captured by our framework, which we discuss in \S \ref{sec:discussion}.  We instead aim to present a reproducible modeling framework that approaches the problem with robust statistics and provides utility in future GO cycles even after the DDT campaign has concluded, as well as in target selection problems in other applications.

This paper is structured as follows: Section \ref{sec:methods} describes our population-level modeling and target selection framework, detailing each modeling step; Section \ref{sec:result_occurence} presents the results for the first experiment and establishes that JWST can efficiently constrain the occurrence rate of atmospheres without optimized target selection; Section \ref{sec:result_cosho} presents the results for the second experiment and demonstrates that the Cosmic Shoreline hypothesis can be distinguished if optimal target selection strategies are employed; Section \ref{sec:discussion} addresses further considerations to be taken into account for choosing targets in addition to the results of the current study.  Our conclusions are summarized in Section \ref{sec:conclusion}.

\section{Methods} \label{sec:methods}

\begin{figure*}[ht!]
    \centering
    \includegraphics[width=0.85\textwidth]{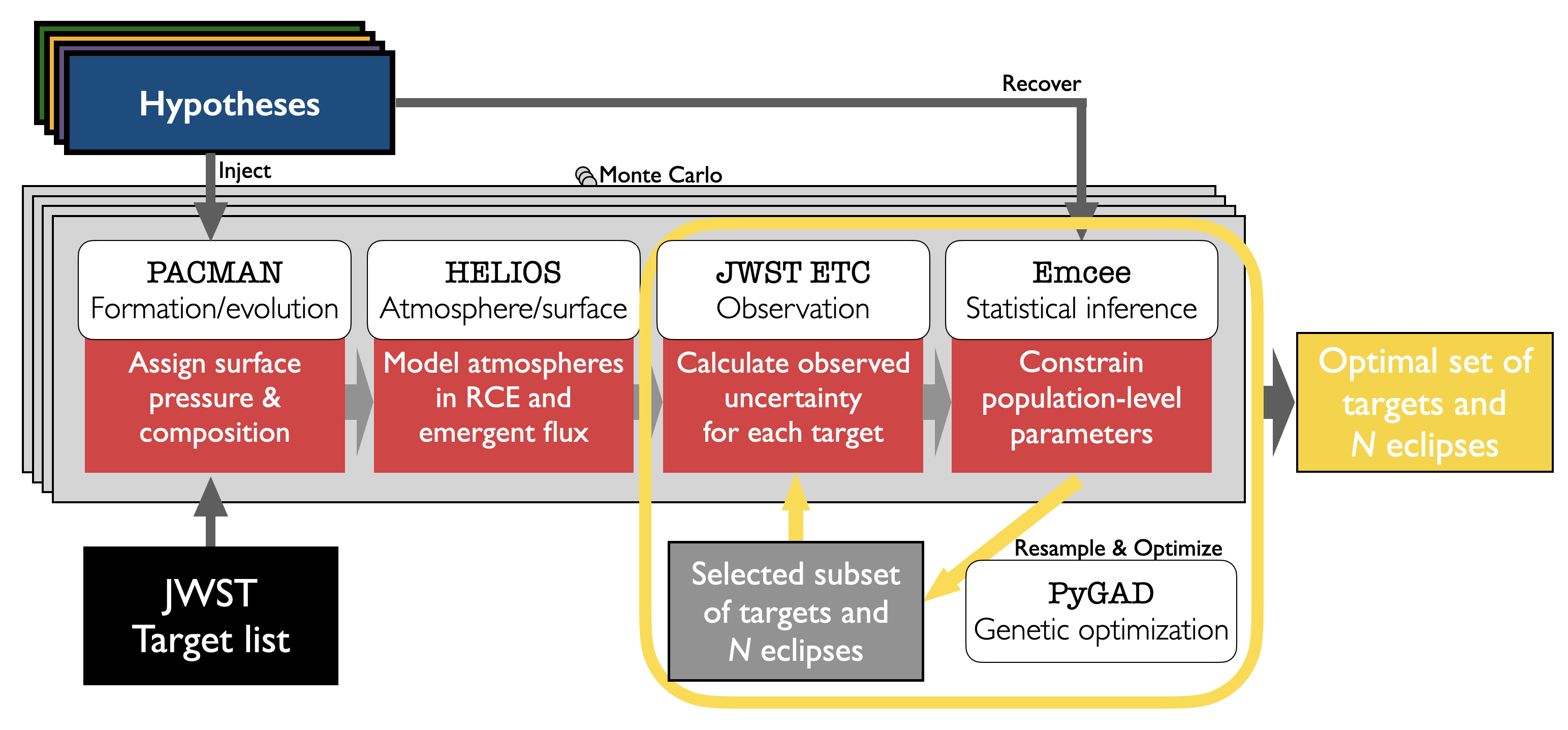}
    \caption{Summary of the population-level injection-recovery simulation.  The pessimist hypothesis is injected for constraining occurrence rate, and the Cosmic Shoreline with and without late M star cut is injected for the hypothesis testing.  Optimal target selection, highlighted in yellow, is performed for the latter only. }
    \label{fig:method_summary}
\end{figure*}


We perform a population-level injection-recovery simulation.  A summary of our methods is shown in Figure \ref{fig:method_summary} as a flowchart.  We present the four parametric hypotheses in \S \ref{subsec:hypotheses}.  We start with a list of targets smaller than 1.7 $R_\oplus$ to be observed with JWST and consider lists constructed via the emission spectroscopy metric (ESM) \citep{kempton18_esm} and the Priority Metric, which we describe in \ref{subsec:targets}.  For each target, we then assign an atmospheric composition and surface pressure sampled from a distribution generated by a formation and evolution model (\S \ref{subsec:pacman}).  We then use \textsc{helios}, a self-consistent radiative-convective equilibrium code \citep{malik17, malik19, whittaker22}, to calculate the one-dimensional thermal profile and its emergent secondary eclipse spectrum for each target (\S \ref{subsec:helios}).  We then use \textsc{pandeia} \citep{pontoppidan16_pandeia} obtain the observed uncertainties for each target(\S \ref{subsec:pandeia}).  We then calculate the inferred probability of an atmosphere, $q_i = prob(\mathrm{atmo})$ for each target from its observed flux by comparing the observation to the depths of the maximally hot blackbody and a 0.1 bar CO$_2$ atmosphere, a choice we justify in \S \ref{subsec:probatmo}.  From the list of calculated $q_i$, we then use Markov chain Monte Carlo (MCMC) to estimate the associated parameters of each hypothesis and its Bayesian information criterion (BIC).  Finally, we use a genetic algorithm implemented in \textsc{pygad} to perform a discrete optimization to select the best set of targets and number of eclipses that maximize the expected difference in BIC between hypotheses (\S \ref{subsec:genetic}).

\subsection{Injected hypotheses} \label{subsec:hypotheses}

To develop testable hypotheses, we must first clearly define the possible scenarios.  We consider four parametric hypotheses, pairs of which we inject and recover:

\textbf{a. Pessimist hypothesis.}  No target has an atmosphere.  A single detection of an atmosphere rejects this hypothesis and answers the question, \textit{does at least one M-star rocky planet have an atmosphere?}

We use three values for how bright the surfaces of \textit{all} planets are: one in which all bare rock planets have blackbody surfaces and one in which they have Bond albedos of 0.2 or 0.4.  The latter is an end member case in which all planets have bright surfaces, which can produce shallow eclipse depths and therefore can act as false positives for the presence of an atmosphere.  We do not model a population for which the surface albedo depends on any specific parameter axes, but discuss the implications in \S \ref{pop_fp}.  This Bond albedo is not estimated when the Pessimist hypothesis is recovered and is fixed to be 0. 

\textbf{b. Random hypothesis.}\footnote[1]{The Pessimist hypothesis, being a point hypothesis, aligns more closely with the common notion of a null hypothesis that can be statistically rejected in a classical hypothesis testing.  In contrast, the Random hypothesis serves as a \textit{baseline} model in which atmospheric occurrence is unstructured across the population. While it cannot be formally rejected in the same sense (since one can always fit an optimal value of $p_\mathrm{int}$), it provides a more useful reference model for comparing structured, physically motivated alternatives.  Within our Bayesian framework, we adopt the Random hypothesis as the effective ``null'' model.}  Targets have atmospheres at random, independent of any specific axes, following an intrinsic binomial probability $p_\mathrm{int}$ between 0 and 1.  Placing an upper limit to this number corresponds to answering the question, \textit{can we statistically conclude if all M-star rocky planets are bare rocks}?

\textbf{c. Cosmic Shoreline hypothesis.} Targets without atmospheres and those likely to have atmospheres are separated by a power law in the (log-)escape speed-bolometric instellation plane with slope $m$ and intercept $I_0$.  Each target is on the lower-right hand side of this line, or the ``wet'' side of the Shoreline, \textit{i.e.} those with $\log{I} < m\log{v_\mathrm{esc}}+\log{I_0}$, with probability $p_\mathrm{wet}$ calculated from the uncertainty in their escape speed (but not instellation).  Targets on the ``wet'' side can have atmospheres with a fixed probability $p_\mathrm{cs}$, while those on the ``dry'' side cannot.   We introduce this probability $p_\mathrm{cs}$ to account for the heterogeneity of M star systems; this turns the Cosmic Shoreline into a permissive hypothesis in which targets on the wet side \textit{may} have atmospheres but not definitively so. Statistically distinguishing between this hypothesis and the Random hypothesis answers the question, \textit{is there a trend in the $V_\mathrm{esc}$-$I$ plane in which M-star rocky planets are likely to have atmospheres?} 

\begin{figure*} \centering
    \includegraphics[width=0.8\textwidth]{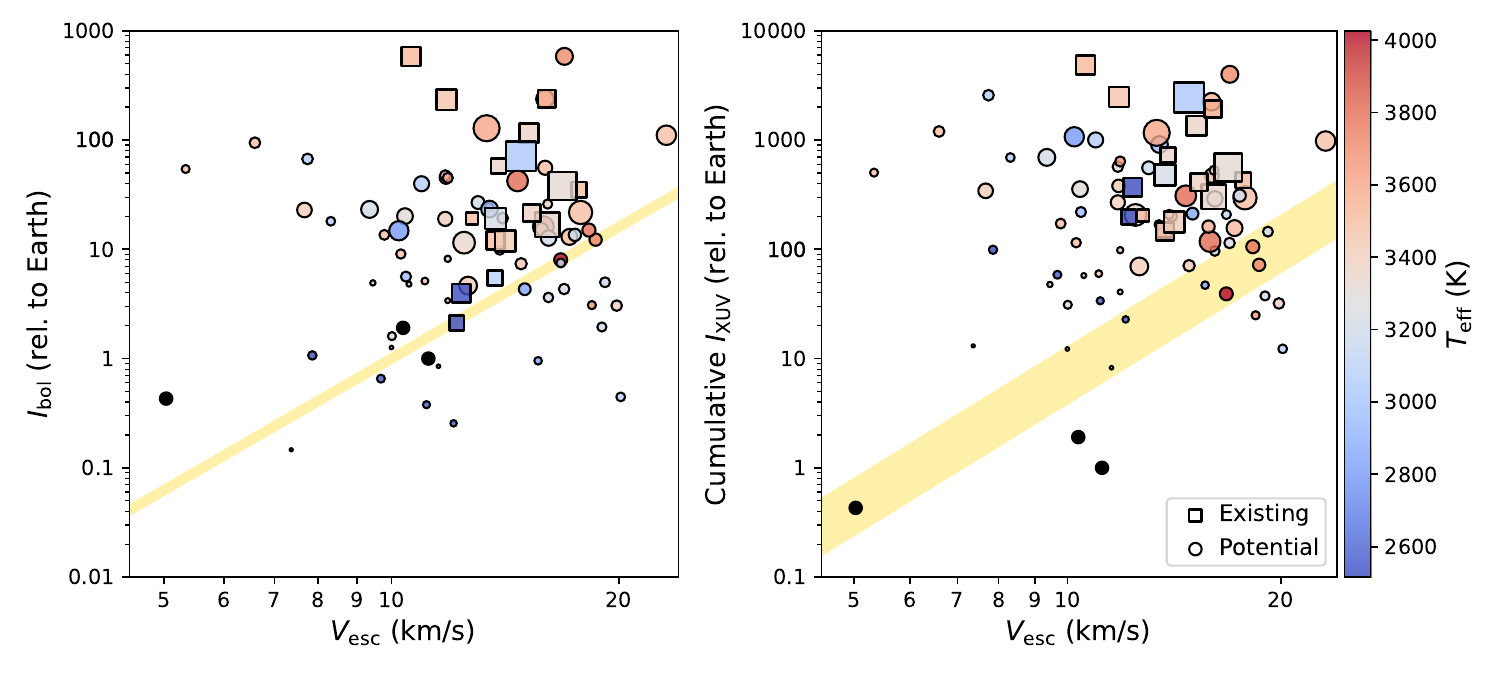}
    \caption{Potential targets and targets already observed plotted in the escape speed-instellation space, in both the instantaneous bolometric flux and the estimated cumulative XUV flux.  The targets are chosen from the DDT ``Targets Under Consideration'' list, which comprises 80 planets.  The color indicates the effective temperature of the host star.  The sizes correspond to the ESM$_{15}$ for each target.  The injected Cosmic Shorelines are also plotted in yellow, drawn to be as low as allowed by existing observations or the Solar system.  The Solar system planets are plotted in black.  The estimated cumulative XUV flux uses the scaling from \citet{zahnle17}, which agrees with stellar age-based estimates to within a factor of $\sim$3 \citep{coy24, pass25}; the vertical width of the line reflects this uncertainty.  }
    \label{fig:cosho}
\end{figure*}

In the injected model, we fix the slope to $m=4$ as per the Solar System value and the intercept to be the most optimistic value allowed by current observations or by the Solar System.  For the bolometric, we choose $I_0 = 10^{-4}$, which passes between TRAPPIST-1 b and c (Fig \ref{fig:cosho}).   We note that this value of $I_0$ places the M-star Shoreline roughly 0.5-dex below the Shoreline in the Solar System; in other words, the assumed Cosmic Shoreline is already informed by the observations to be more pessimistic than one fit only to the Solar System.    Also, we do not consider a situation in which $p_\mathrm{cs}$ is a function of the target's distance from the Cosmic Shoreline (CPM), but do acknowledge that this is a possibility \citep{berta25}.

\textbf{d. XUV Cosmic Shoreline hypothesis}.  Same as previous, but instead of current bolometric flux, we use cumulative XUV flux to determine whether a target can have atmospheres.  We use the scaling from \citet{zahnle17} to calculate the cumulative XUV flux given a current (bolometric) instellation and the stellar effective temperature. Statistically distinguishing this hypothesis from the original Cosmic Shoreline hypothesis answers the question, \textit{does the M-star Cosmic Shoreline differ from the Solar System Shoreline?}

In the injected model, we choose the intercept to be $I_0 = 10^{-3}$; the value that passes through Mars as drawn in Fig \ref{fig:cosho} is $I_0 = 7 \times 10^{-4}$, but we adopt a greater value to account for the upper limit in the uncertainty in cumulative XUV scaling (as indicated by the width of the line in Fig \ref{fig:cosho}).



During recovery, the four hypotheses have $k=0$, 1, 3, and 3 free parameters, respectively, as shown in Table \ref{tab:params}.  In the recovery of the Cosmic Shoreline hypotheses, we allow the slope of the line to be a free parameter and not fixed to 4, as the Cosmic Shoreline is empirically motivated.

Using the hypotheses, we perform two experiments: 

\textbf{Experiment \#1}: We inject the Pessimist hypothesis and recover the Random hypothesis to place a constraint on the occurrence rate of atmospheres \textit{if planets simply had atmospheres at random}, $p_\mathrm{int}$.  We repeat for injected values of $A_\mathrm{B}=0.2, 0.4$ to test for robustness against bright surfaces and also test a number of priors on $p_\mathrm{int}$. 

\textbf{Experiment \#2}: We inject the Cosmic Shoreline hypotheses and recover the injected Cosmic Shoreline hypothesis and the Random (or Pessimistic) hypothesis, then find the maximal value of $\mathbb{E}[\Delta \mathrm{BIC}]$ between the two hypotheses by optimizing for the set of observations.  We do this for both the bolometric and the XUV versions of the Cosmic Shorelines.  We also repeat this for the Pessimist hypothesis in place of the Random hypothesis as the null hypothesis.

\begin{table*}[]
    \centering
    \begin{tabular}{c|c|p{7cm}|c}
        \textbf{Hypothesis} &  \textbf{Parameter} & \textbf{Description} & Injected values  \\ 
        \hline \hline
          \textbf{a. Pessimist}  &  $A_\mathrm{B}$ & Bond albedo for all bare rock planets (not recovered)  & 0, 0.2, 0.4 \\ \hline
         \textbf{b. Random}       & $p_\mathrm{int}$ & Probability of any planet having an atmosphere  & - \\ \hline
        \textbf{c. Bol. Cosmic Shoreline} & $m$ & Slope of the Cosmic Shoreline (CS) in $\log{v_\mathrm{esc}}$-$\log{I_\mathrm{bol}}$ plane & 4 \\
        & $\log{I_0}$ & Log-intercept of the CS & -4 \\
        & $p_\mathrm{cs}$ & Probability of a planet on the wet side of the CS having an atmosphere & 0.33, 0.50, 1.00 \\ \hline
         \textbf{d. XUV Cosmic Shoreline} & $m$ & Slope of the CS in the $\log{v_\mathrm{esc}}$-$\log{I_\mathrm{xuv}}$ plane & 4 \\
        & $\log{I_0}$ & Log-intercept of the XUV CS & -3 \\
        & $p_\mathrm{cs}$ & Probability of a planet on the wet side of the XUV CS having an atmosphere & 0.33, 0.50, 1.00 \\ 
    \end{tabular}
    \caption{Summary of free parameters for each hypothesis.  The Pessimist hypothesis has no free parameters when it is being recovered.}
    \label{tab:params}
\end{table*}


\subsection{Best-in-class targets for emission observations} \label{subsec:targets}

\begin{figure*}[!ht]
    \includegraphics[width=\textwidth]{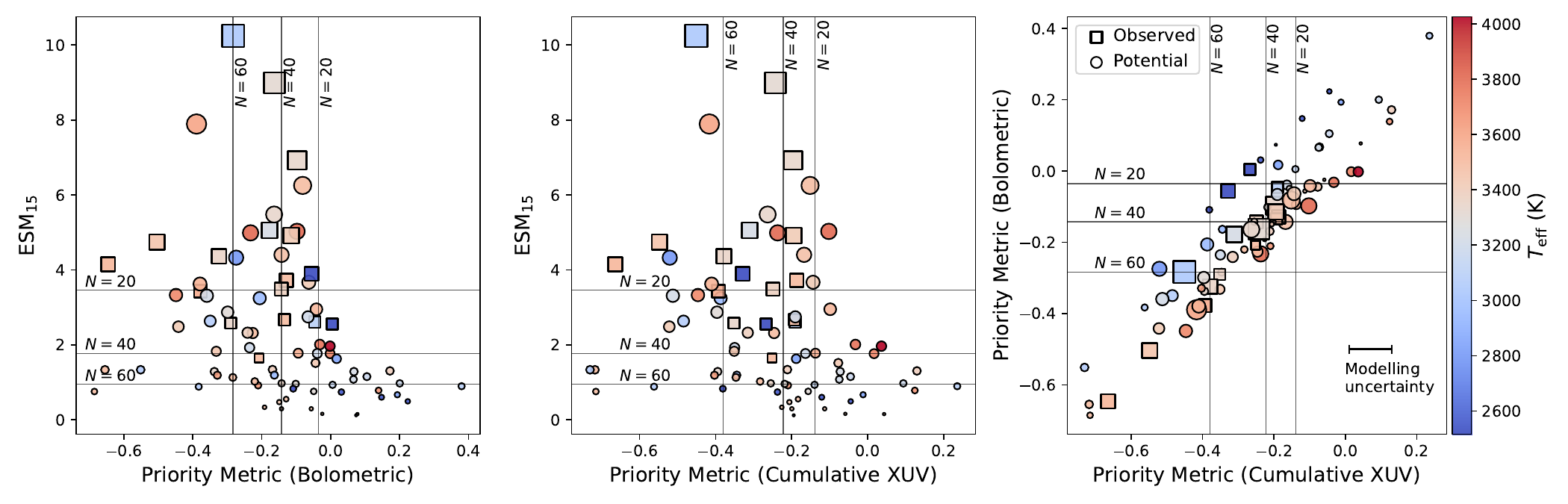}
    \caption{The 80 targets ranked by ESM$_{15}$ and the bolometric Priority Metric (\textit{left}), by ESM$_{15}$ and the cumulative XUV Priority Metric (\textit{center}), and by the two Priority Metrics.  The targets are chosen from the DDT ``Targets Under Consideration'' list.   The vertical and horizontal lines indicate quartiles along each dimension.  The color indicates the effective temperature of the host star, with red (blue) markers correspond to targets orbiting early (late) M stars.  The targets are scaled by their ESM$_{15}$.  The factor of $\sim 3$ disagreement in cumulative XUV between the scaling in \citet{zahnle17} and age-based calculations \citep{coy24}, propagated to the priority metric, is shown as the modeling uncertainty in the right panel; this should not be construed as a 1-$\sigma$ uncertainty.}  
    \label{fig:pm_esm}
\end{figure*}

The targets are chosen from the DDT ``Targets Under Consideration'' list, which comprises 80 planets \footnote[2]{https://rockyworlds.stsci.edu/rw-website-targets.html}.   

We note, importantly, that not all targets in the list have known masses; for these targets,  masses have been estimated from the radius using the \textsc{sprite} code \citep{parviainen23_spright}, which takes a prior on density based on a prior on iron-to-rock ratios.  Additionally, three of the targets included in the list (LHS-1140 b, TOI-406.01 and TOI-5388.01) have radii greater than 1.7 $R_\oplus$, but all of them have uncertainties large enough to be consistent with this radius at 3 $\sigma$ \citep[e.g.][]{luque24_dark}, so we include them in the sample \citep{cadieux24_lhs1140b,  hord24_toi406, hord24_identification} .  We accept the values for planet radius and mass as point values, but discuss the potential pitfalls of ignoring error bars and the need for precise mass measurements in \S \ref{subsec:mass}. 

Target lists ranked by some metric provide useful benchmarks for target selection.  We consider two lists constructed via rank ordering of the targets based on two metrics: one via the emission spectroscopy metric (ESM) \citep{kempton18_esm}, as done in \citet{hord24_identification}, and another via the ``Priority Metric'', which we describe below. 

We modify the definition of the ESM to be more suited to MIRI photometry as follows.  Firstly, we calculate ESM$_{15}$ based on the blackbody flux referenced at 15~$\mu$m, which is more suitable to observing rocky planets, whereas \citet{kempton18_esm} used 7.5 $\mu$m assuming LRS observations.  Secondly, we use the maximally hot temperature, or $T_\mathrm{max}$, \textit{i.e.} that given by a bare blackbody, to calculate the ESM rather than using $1.1$ times the equilibrium temperature, which was motivated by predictions from global circulation models assuming the targets had at least tenuous atmospheres.  Finally, we choose the leading scaling factor $A$ to match the empirical signal-to-noise of a single eclipse TRAPPIST-1 b reported in \citet{greene23}, such that ESM$_{15}$ roughly matches the expected signal-to-noise for a single eclipse.  In short, we define ESM$_{15}$ as:

\begin{equation}
    \mathrm{ESM}_{15} = A \frac{B_\lambda(T_{max}; 15\mu \mathrm{m})}{B_\lambda(T_{star}; 15\mu  \mathrm{m})} \frac{R^2_\mathrm{p}}{R^2_\mathrm{s}} 10^{-m_K / 5},
\end{equation}

where $A=7.2 \times 10^5$ is the chosen scaling factor, $B_\lambda(T;\lambda)$ is the Planck function, and $m_K$ is the K-band magnitude.  We note that the brightness temperature of a star may significantly differ from its effective temperature at 15 micron, as much as $\sim 30 \%$ \citep{fauchez25}, and using a blackbody is an approximation; similarly for using readily available K-band magnitude than the actual 15-micron magnitudes.  However, as ESM$_{15}$ already has an uncertainty propagated from the the orbital distance, planet radius, and host star radius, we deem the approximation and empirical scaling to TRAPPIST-1 b good enough.   We also note that the ordering of targets by ESM and ESM$_{15}$ loosely correspond to each other but are not identical.

 
Additionally, for the target selection problem, we also consider a target list chosen based on rank ordering by the Priority Metric, defined as the orthogonal distance to the Cosmic Shoreline, or:

\begin{equation}
    \mathrm{PM} = (4 \log_{10}(V_\mathrm{esc}) - \log_{10}(I) + \log_{10}(I_0)) / \sqrt17,
\end{equation}

where $I$ can refer to either the bolometric or the cumulative XUV flux.  This value depends on the assumed intercept of the Cosmic Shoreline and therefore is free up to a constant; a Priority Metric of zero depends on the assumed intercept. 

We plot the resulting targets rank ordered by each metric in Figure \ref{fig:pm_esm}, where we also show the quartiles.  There is little or no overlap between the best 20 ESM$_{15}$ targets and the best 20 Priority Metric target; zero overlapping targets for bolometric flux metric and one overlapping target (HD 260655 c) for cumulative XUV metric.  This is unsurprising given that ESM$_{15}$ favors the hottest targets, while the Priority Metric favors the lowest instellation and therefore the coolest targets.

The choice between bolometric and cumulative XUV Priority Metric entails some subtlety.  There are 15 overlapping targets in the best 20 Priority Metric targets; the remaining choice is between early M stars, favored by the cumulative XUV Priority Metric, shown in blue, and late M stars, favored by the bolometric Priority Metric, shown in red.  Given that (a) there are non-negligible uncertainties in estimating the cumulative XUV flux, indicated by the horizontal bar, and (b) the best Priority Metric targets also tend to be the lowest ESM$_{15}$ and hence the most expensive targets, we cannot simply rank order by the Priority Metric to find the most efficient target selection strategy.  Regardless, the Priority metric remains a notionally useful number to indicate which targets are the closest to the Cosmic Shoreline.

\subsection{Formation \& Evolution model} \label{subsec:pacman}

From the given injected hypothesis, we draw a random instance of a population that varies in whether each potential target has an atmosphere or not.  Based on this binary distinction, we use outputs from \textsc{pacman}, a coupled interior-atmosphere evolution model \citep{krissansen21, krissansen22_predictions}, to obtain a realistic distribution of the atmospheric compositions and the total surface pressures of the atmosphere.  Simulating formation and evolution for every single target is computationally prohibitive due to the large number of physical parameters.   As a simpler alternative, we sample from the distribution of outputs from \citet{krissansen22_predictions}, originally generated for the TRAPPIST-1 planets where the distribution of compositions were bootstrapped from the prior distribution of unconstrained physical parameters.   As the H$_2$O abundance can vary based on surface condensation, we make a distinction between planets interior to and within the habitable zone.  We define the habitable zone simply as those receiving less bolometric flux than the Earth \citep{kasting93, kopparapu13}.  We sample compositions from the distribution for planet b for the former and e for the latter---but note that only a handful (7 out of 80) of targets reside within the habitable zone.

We deem this simplification good enough as the exact composition of the atmosphere acts as a nuisance parameter that we ultimately wish to marginalize over by Monte Carlo sampling.  To first order, the  observationally relevant output is the partial pressure of CO$_2$ and to a lesser extent H$_2$O \citep{malik19b, ih23}.  The \textsc{pacman} model captures the plausible range of the abundances of these gases, especially their ubiquity, which we comment on in \S \ref{subsec:co2}.

\subsection{Atmosphere model} \label{subsec:helios}

Once we have an assigned atmospheric composition and total surface pressure for each potential target, we use the open-source code \textsc{helios} to model the thermal structure and the emergent flux of the planet \citep{malik17,malik19,whittaker22}.  We include the same sources of gas absorption and scattering, as well as collision-induced absorption, as used in \citet{whittaker22}.  We assume a blackbody surface at the bottom of the atmosphere in radiative equilibrium wiht the overlying atmosphere.  For the input stellar spectrum, we use the \textsc{sphinx} model grid and interpolate to the point values of the stellar parameters, with the interpolation for the effective temperature done in log scale \citep{iyer23,Wachiraphan24}.  The \textsc{sphinx} model originates from the \textsc{phoenix}/\textsc{BT-Settl} family of stellar models and has improvements specific to lower mass stars, such as updated molecular line list sources and has shown to be better at fitting observed stellar fluxes \citep{fauchez25}.

We make the caveat as for any simulation study that using the same model for both forward and inverse modeling is almost guaranteed to be somewhat optimistic in what constraints are possible \citep{whittaker22}.  Specifically to simulations of M-star rocky planets, this is perhaps most true in the uncertainty in the stellar model, which have at times struggled to match observations \citep{ih23, coy24, fauchez25}.  For modeling the planet, the forward model necessarily makes assumptions such as a one-dimensional vertical atmosphere and parameterized heat redistribution; one can only surmise that, in reality, there are subtle missing physics that act as systematics of physical origin and produce measurable consequences.  We leave cross-validation between different models and testing which assumptions are consequential for future work.

\subsection{Observation model} \label{subsec:pandeia}

We use the \textsc{pandeia-engine v4.0} to estimate the uncertainty in the eclipse depth of a single eclipse for each target.  We assume that the targets have circular orbits, and therefore transit duration and eclipse duration are equivalent (we discuss the consequences of possible deviations in \S \ref{sec:discussion}.  For speed of calculation and being agnostic to the dayside temperature, we assume that to first order the eclipse depth error can be estimated from the ETC for a single integration, and then binned down given the number of integrations during the eclipse duration.  We do not explicitly include the effect of time-dependent systematics that can appear in MIRI imaging \citep[e.g.,][]{august2024}, which we discuss in \S \ref{sec:discussion}; however, where we require a signal-to-noise limit we set a minimum of 4 $\sigma$ distinction (rather than the more conventional 3 $\sigma$) to account for this, effectively inflating the errors by 33\%. 

To estimate the observing time of each target, we adopt an out-of-eclipse baseline time equal to the eclipse duration or round up to 1 hour if the eclipse duration is shorter than 1 hour, following standard practices in the field \citep{hotrocks}.  We add 1.5 hours to the total time to account for the flexible observing start time and detector settling.  Finally, to estimate the charged time, we apply a flat 40\% overhead to the calculated science time incurred by using JWST.  The exact overhead will vary slightly (of order a few percent) for each target and each observation; but 40\% is a reasonable approximation based on e.g., the \textit{Hot Rocks Survey}.

\subsection{Probability of atmosphere for eclipse observations} \label{subsec:probatmo}

Once we have a randomly drawn observation of eclipse depth $d_i$ and uncertainty $\sigma_i$ for each target, we then assign a probability of it having an atmosphere, $q_i$.  The exact method of how to assign this number is predicated on what it means to ``have an atmosphere'', which is, of course, a subjective and fuzzy choice.  Here, we provide our definition and justify our choice.

For each target, we consider two mutually exclusive possibilities that the $i$-th planet ``has an atmosphere'' $\mathcal{A}_i$ and that it ``is a bare rock'' $\mathcal{B}_i$.  We define $q_i$ as the probability of the former being true given an observation, or $q_i = p(\mathcal{A}_i \vert d_i)$.   The prior and posterior possibilities sum to 1, \textit{i.e.} $p(\mathcal{A}_i)+p(\mathcal{B}_i)=1$ and $p(\mathcal{A}_i \vert d_i)+p(\mathcal{B}_i \vert d_i)=1$. 

From this, we can write $q_i$ as a posterior odds ratio in terms of the more familiar Bayes factor $B_{\mathcal{AB},i}$:

\begin{equation}
    q_i = \frac{B_{\mathcal{AB},i}}{1+B_{\mathcal{AB},i}},
\end{equation}

\noindent where the Bayes factor is calculated as:

\begin{equation}
    B_{\mathcal{AB},i} = \frac{p(d_i \vert \mathcal{A}_i)}{p(d_i \vert \mathcal{B}_i)} = \frac{\int p(d_i \vert \boldsymbol{\theta}_\mathcal{A}) \pi(\boldsymbol{\theta}_\mathcal{A} \vert \mathcal{A}_i ) d \boldsymbol{\theta}_\mathcal{A}} {\int p(d_i \vert \boldsymbol{\theta}_\mathcal{B}) \pi(\boldsymbol{\theta}_\mathcal{B} \vert \mathcal{B}_i ) d \boldsymbol{\theta}_\mathcal{B}},
\end{equation}

\noindent where $\pi(\boldsymbol{\theta}_\mathcal{A})$ and $\pi(\boldsymbol{\theta}_\mathcal{B})$ indicate the priors on parameters associated with a planet having an atmosphere and being a bare rock, respectively, such as surface pressure, gas composition, clouds, bond albedo, or surface mineralogy.  

One could, in principle, select sensible priors for such parameters and use a retrieval model to calculate the integrals.  This step would necessarily reflect one's subjective choice, \textit{e.g.} below what surface pressure a planet no longer ``has an atmosphere'', which clouds are plausible, or how bright an albedo a bare rock planet can truly have.  However, we do not have an obvious prior for these parameters.  More importantly, a retrieval would take much longer time than is required to make numerous evaluations of $q_i$ feasible.

As such, we instead use a point estimate in the limit using a canonical ``shallow-eclipse'' model of 0.1 bar CO$_2$ atmosphere as a proxy for $\mathcal{A}$ and and a maximally hot blackbody surface for $\mathcal{B}$:

\begin{equation}
    B_{\mathcal{AB},i} \approx \frac{p(d_i \vert \boldsymbol{\theta}_{\mathcal{A}_0} ,  \mathcal{A}_i)}{p(d_i \vert \boldsymbol{\theta}_{\mathcal{B}_0} , \mathcal{B}_i)} = \frac{\mathcal{L}_{\mathrm{shallow} ,i}}{\mathcal{L}_{\mathrm{bare}, i}},
\end{equation}

\noindent where the likelihoods $\mathcal{L}$ for shallow eclipse depth and a bare rock follow the usual normal distribution:

\begin{equation}
    \ln\mathcal{L}_{\mathrm{bare},i} = -\frac{1}{2} (d_i - d_{\mathrm{bare},i})^2 / \sigma_i^2 + \ln2\pi\sigma_i^2,
\end{equation}

\noindent and similarly for $\mathcal{L}_{\mathrm{shallow}}$, where $\sigma$ is the binned error after $N$ eclipses for each target, assumed to scale as $1/\sqrt{N}$.  This assumes that we can simply stack eclipses and bin down the errors, an assumption likely to be somewhat optimistic given the presence of instrumental systematics.  We discuss the validity of this assumption in \S \ref{subsec:systematics}.  The likelihood is defined for a single photometric channel, but can be easily generalized to a multi-band or spectral observation as the product of each likelihood, assuming independent errors. 

Then, $q_i$ is simply the relative likelihood ratio between a 0.1 bar CO$_2$ atmosphere and a maximally hot blackbody surface:

\begin{equation}
    q_i = p(\mathcal{A}_i \vert d_i) = \frac{1}{1 + \mathcal{L}_{\mathrm{bare},i} / \mathcal{L}_{\mathrm{shallow},i}}. 
    \label{eqn:prob}
\end{equation}

\begin{figure}
    \includegraphics[width=\columnwidth]{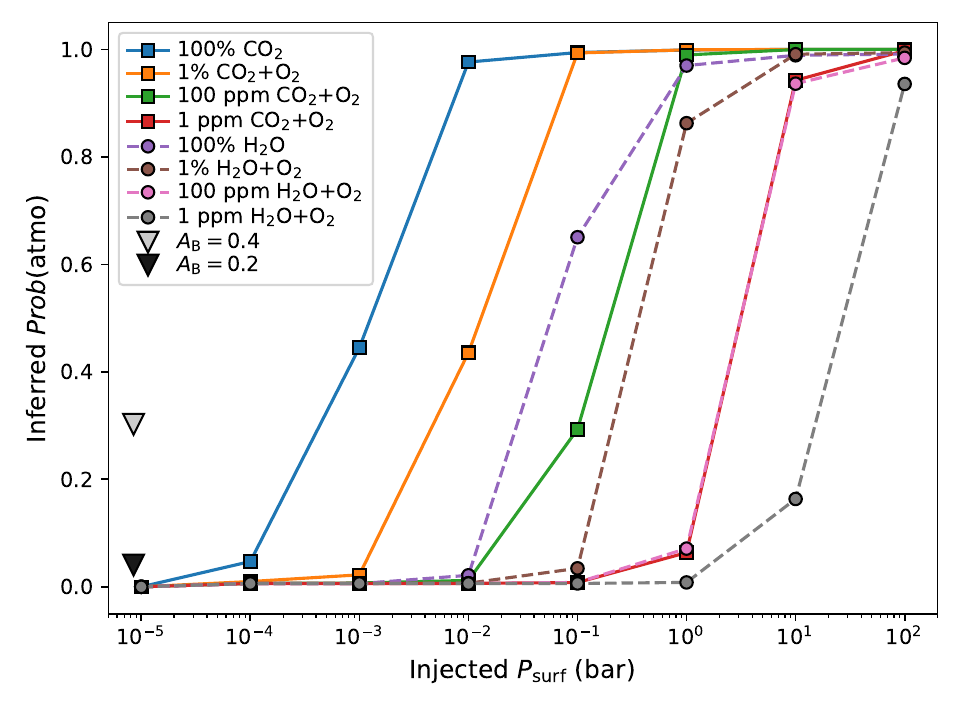}
    \caption{The inferred probability of an atmosphere as defined in Equation \ref{eqn:prob} as a function of the actual modeled surface pressures for a number of simulated end member compositions.  The probability is calculated for a single F1500W eclipse observation of GJ 1132 b.}
    \label{fig:probatmo}
\end{figure}

To be explicit about our priors indicated by the point estimates using 0.1 bar CO$_2$ atmospheres as the canonical model, all inferred $q_i$ assumes that if a \textit{detectable} atmosphere exists, it \textit{looks like} 0.1 bar CO$_2$, even though real atmospheres could be thicker/thinner or composed of different gases.  We discuss the implications for the population-level inference in \S \ref{subsec:co2}.  For a range of plausible compositions considered, the eclipse depth of the 0.1 CO$_2$ model is a good approximation for the vast majority of our models that possess atmospheres, since most models include some CO$_2$ and redistribute heat, resulting in shallower than bare-rock eclipses.  We plot the inferred $q_i$ values for a single eclipse of sample planet GJ 1132 b for a grid of compositions, surface pressure, and Bond albedos in Fig \ref{fig:probatmo}.   Increasing the number of eclipses has the effect of sharpening the sigmoidal function closer to a step function between $q_i$ of 0 and 1, but does not alter where the shift occurs.



\subsection{Population-level statistical inference} \label{subsec:pop}


Once we have the list of targets and their calculated values of $q_i$'s, we estimate the parameters associated with each hypothesis.  If the full posterior distribution is needed, we use Markov chain Monte Carlo (MCMC) implemented in \texttt{emcee} \citep{foreman13} to draw posterior samples.  If only the Bayesian information criterion (BIC) is needed, we use the Powell optimizer as implemented in \texttt{scipy.optimize} to find the parameters values that maximize average BIC for a given hypothesis for each realization.  The difference $\Delta$BIC, averaged over different realizations, is then used as the fitness function for the genetic algorithm to optimize the list of targets over (\S \ref{subsec:genetic}). 


BIC is a point estimate for evidence defined using the maximum likelihood for a given hypothesis \citep{raftery1995}.  For a given set of observations $\{d_i\}$, the population-level likelihood $\mathcal{L}_{\mathrm{pop}}$ for hypothesis $\mathcal{H}$ and its parameters $\boldsymbol{\Theta}$ is defined as:

\begin{equation}
    \begin{aligned}
    \mathcal{L}_{\mathrm{pop}} &= p(\{d_i\} \vert \boldsymbol{\Theta}, \mathcal{H}) \\ &= \prod_i p(d_i \vert \mathcal{A}_{i}) p (\mathcal{A}_{i} \vert \boldsymbol{\Theta}, \mathcal{H})+ p(d_i \vert \mathcal{B}_{i}) p (\mathcal{B}_{i} \vert \boldsymbol{\Theta}, \mathcal{H}), 
    \end{aligned}
\end{equation}

\noindent which can be rewritten using $q_i$ as: 

\begin{equation}
    \mathcal{L}_{\mathrm{pop}} =\prod_i [q_i p_{i} + (1 - q_i) ( 1- p_{i})](\mathcal{L}_{\mathrm{shallow},i} + \mathcal{L}_{\mathrm{bare},i}),
\end{equation}

\noindent where $p_{i} = p(\mathcal{A}_{i} \vert\boldsymbol{\Theta}, \mathcal{H})$ is the probability the $i$-th target should have an atmosphere given the chosen hypothesis $\mathcal{H}$ and the sampled population-level parameters.  The term in the square brackets reflects the probability that the sampled hypothesis correctly predicts whether a given target has an atmosphere, while the second term is the evidence.  For testing the Random hypothesis, $p_i$ is fixed for all targets to be $p_i = p_\mathrm{int}$;  for the Cosmic Shoreline hypothesis, $p_i=p_\mathrm{wet} \, p_\mathrm{cs}$, wherein $p_\mathrm{wet}$ is the probability of a target being on the wet side of the Cosmic Shoreline based on its uncertainty from $v_\mathrm{esc}$.  

From the maximum likelihood estimation, we calculate the BIC for each hypothesis and the difference $\Delta \mathrm{BIC}$ among them.  BIC is defined as $\mathrm{BIC} = k\log{N} - 2 \log{\mathrm{max}(\mathcal{L}_{\mathrm{pop}})}$, where $k$ is the number of parameters for each hypothesis and $N$ is the number of planets included in each candidate target list.  For the pessimist hypothesis, which has no free parameters, BIC is simply equal to $-2 \log{\mathcal{L}}$.  One important feature of BIC is that it does not explicitly depend on the priors on $\boldsymbol{\Theta}$, as it already assumes weakly informative priors centered around the maximum likelihood estimate \citep{kass95}, which one would generally expect to be the case for injection-recovery simulations.  In the limit where priors are uninformative and $N \gg k=\mathrm{const}.$ (\textit{i.e.} BIC$\gg1$), the BIC approximates (two times) the Bayes factor between the two hypotheses \citep{raftery1995}.   


It is consequential that we maximize the \textit{difference} in BIC between two hypotheses, $\Delta$BIC, rather than simply minimizing the BIC of one given hypothesis.  Like the Bayes factor, BIC only has significance in relative terms; also, BIC is suited to select between different models, not to select between different samples.  For instance, the optimizer could minimize the BIC for Cosmic Shoreline hypothesis by picking only the bare rock targets and drawing a line below all selected targets.  However, in order to simultaneously maximize the BIC for the Random hypothesis, the optimizer must pick targets on both sides of the Shoreline to select the set of targets that are the most incongruent with being sampled from a uniform $p_\mathrm{int}$.  

\subsection{Target selection using a genetic algorithm}  \label{subsec:genetic}

For the optimal target selection, we apply a genetic algorithm (GA) implemented in the \textsc{pyGAD} package \citep{gad2023pygad}.  A GA is an optimization method loosely based on biological evolutionary principles of random mutation and natural selection \citep{charbonneau95_genetic}.  A GA is well suited to solve a knapsack problem, \textit{i.e.} a combinatorial optimization problem, when the problem is highly non-linear and potentially multi-modal, as GA is much less likely to be trapped in local maxima than local approaches \citep{ford05}.

To briefly describe how a GA works, a GA begins with an initial population of candidate solutions. In our application, each solution is a potential survey that consists of 80 genes that each encode the number of eclipses for each target (which may be zero), wherein the targets are sorted by their ESM$_{15}$.  For each candidate solution, we calculate the fitness function, which we choose to be the expected $\mathbb{E}[\Delta \mathrm{BIC}]$ between a chosen pair of hypotheses (averaged over  multiple realizations of the injected hypothesis).  For a candidate solution whose total amount of observing time exceeds the specified total time of 550 hours (500 hours plus 10\% tolerance), we assign a fitness function of zero.  Then, in each successive generation, the population evolves; each  solution is retained, altered, or replaced with an offspring of solutions with higher fitness by selection, mutation, and crossover operations, respectively, whose parameters are chosen by the user \citep{hassanat19}.  We halt the evolution if the best fitness function in the population does not improve over a specified number of generations.  Then, in the final population, we take the solution with the best fitness function as the optimized set of targets.

A GA does not \textit{guarantee} global optimality.  It is necessary to fine tune the optimization parameters so that we can be practically confident that we have a satisfactory solution.  For this, parameters that sufficiently explore the diversity within the solution space, especially near the total time limit, is important.  We experimented with a number of optimization parameters and found the choice of parameters listed in Table \ref{tab:gad_params} to be adequate.  

The strategy to search for new solutions (exploration strategy) is to allow each gene have a 2\% chance to randomly shift (\texttt{mutation\_percent\_genes}) by as many as 2 eclipses in either direction (\texttt{random\_mutation\_val}) every generation and to select a contiguous subset of genes and shuffle their values randomly every generation (\texttt{mutation\_type}).   The strategy to retain and refine a good candidate solution (exploitation strategy) is to select and keep the best 10 solutions in each population (\texttt{keep\_elitism}). Additionally, for 10\% of the population (\texttt{num\_parents\_mating}), we choose the best out of 5 randomly drawn (\texttt{K\_tournament}) parents, who in turn have 70\% chance of generating an offspring (\texttt{crossover\_probability}), which then replaces a solution of lower fitness.  Finally, we stop the optimization if the best fitness function in the population does not improve over 200 generations (\texttt{stop\_criteria}).   

We find the optimal set of targets that perform the best across Monte Carlo draws \textit{on average}.  Given a candidate solution, we use the expected value, or the arithmetic mean, of $\Delta \mathrm{BIC}$ values as the fitness function.  We note explicitly that this is a choice that we make; one could instead, adopt a min-max strategy and choose the worst-case $\Delta$BIC as a fitness function.  However, as doing so has a potential pitfall of becoming sensitive to outliers, we remain optimistic and adopt the expected value as the fitness function.

\begin{table}[]
    \centering
    \begin{tabular}{l|c}
        \texttt{Parameter} & \quad Value \\\hline
        \texttt{sol\_per\_pop} & \quad 500 \\
        \texttt{num\_genes} & \quad 80 \\
        \texttt{stop\_criteria} & \quad \text{``saturate\_200''} \\
        \texttt{parent\_selection\_type} & \quad \text{``tournament''} \\
        \texttt{num\_parents\_mating} & \quad 50 \\
        \texttt{K\_tournament} & \quad 5 \\
        \texttt{keep\_parents} & \quad 5 \\
        \texttt{keep\_elitism} & \quad 10 \\
        \texttt{crossover\_type} & \quad \text{``two\_points''} \\
        \texttt{crossover\_probability} & \quad  0.7   \\
        \texttt{mutation\_type} & \quad \text{``random'', ``scramble''} \\
        \texttt{mutation\_by\_replacement} & \quad \text{False} \\
        \texttt{random\_mutation\_min\_val} & \quad -2    \\
        \texttt{random\_mutation\_max\_val} & \quad 2 \\
        \texttt{mutation\_percent\_genes} & \quad 2\%
    \end{tabular}
    \caption{The hyperparameters used for optimization via \textsc{pyGAD}.  The values were chosen based on experimentation to find hyperparameters that produce an optimal solution.  The parameters chosen here prioritize a broad enough search to not get stuck in local minima.}
    \label{tab:gad_params}
\end{table}

\section{Results \#1: Measuring the Occurrence Rate of Atmospheres} \label{sec:result_occurence}

In this section, we present the results of the first experiment, in which we quantify what constraints on the occurrence rate of an atmosphere are achievable and decide how many observations of bare rocks are necessary to conclude that the full population consists of only bare rocks.  To do this, we produce lists of varying sample sizes, rank ordered by ESM$_\mathrm{15}$.  We inject the Pessimist hypothesis and recover the Random hypothesis to obtain the posterior distribution on the intrinsic probability of an atmosphere, $p_\mathrm{int}$.  We vary the sample size to test how the posterior on $p_\mathrm{int}$ varies with the sample size and total observing hours.

Our five target lists follow from total observing hours (charged time) limits of 20, 100, 500, and 2500 hours and the full sample of 80 targets.  For each target, we require that we need to stack enough eclipses to distinguish between the eclipse depths that arise from a bare rock and from a 0.1 bar CO$_2$ atmosphere at 4 $\sigma$.  This results in ESM$_\mathrm{15}$ cuts of 7.8, 4.5, 2.66, and 1.33, respectively, yielding 4, 10, 27, and 47 targets in the sample.   In summary, rank ordered by ESM, \textit{doubling} the sample size requires roughly five times more observing hours, as samples included later in the list are of lower signal-to-noise.

We repeat over 50 Monte Carlo draws, wherein only the observational noise instance per target varies in each draw (since all planets are fixed to be bare rocks). While posteriors of binomial distributions can be calculated analytically, the bootstrapping allows for propagating observational uncertainties to the estimated confidence intervals. We find that the population-level posterior is generally not sensitive to the particular Monte Carlo instance in this experiment, as the imposed 4-$\sigma$ threshold safeguards against outliers.

\begin{figure*}[ht!] 
    \centering
    \includegraphics[width=0.8\textwidth]{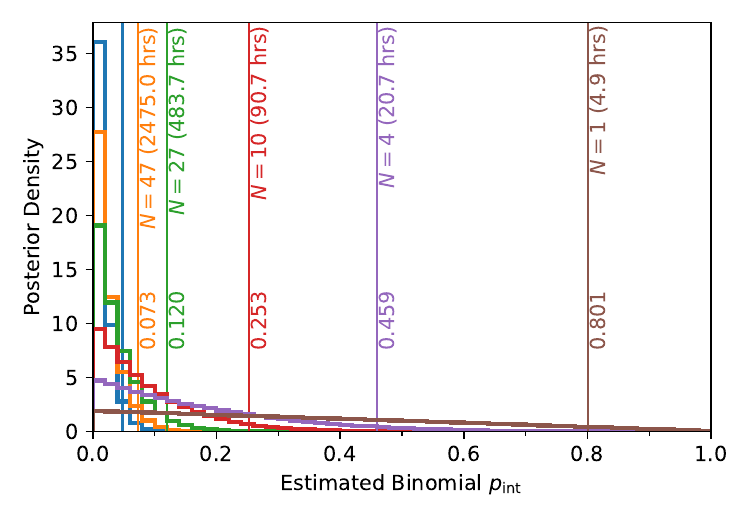}
    \caption{The constraints on measured intrinsic probability of targets having an atmosphere for surveys of different sizes, where all potential targets are bare rocks in the ground truth.  The histograms show the posterior distribution from target lists based on different ESM cuts, as well as the total number of targets and hours.  The vertical  lines show the 95 \% confidence intervals for each target list.   }
    \label{fig:posterior_probatmo}
\end{figure*}

\subsection{Baseline case: blackbody surfaces \& uniform prior}

We present the resulting posteriors for $p_\mathrm{int}$ in Figure \ref{fig:posterior_probatmo}, with the 95\% confidence interval (CI) upper limits indicated by vertical lines.  We assume a uniform prior on $p_\mathrm{int}$.  Based on the four ESM cuts, the 95\%-CI upper limits are 45.9\%, 25.2\%, 11.6\%, 7.1\% for 20.7, 90.7, 483.7, 2470 hours of observing time, respectively.  Observing the maximum hour sample gives a 95\%-CI upper bound of 4.6\%.  Another way to interpret these numbers is the following: if we were to observe 27 planets and find all of them to be bare rocks, the strongest conclusion we could draw is that at most 1 in 10 M-star rocky planets have atmospheres at 95 \% confidence interval. From the Monte Carlo, which captures the observational uncertainty, we determine that these numbers have approximately 2\% uncertainty.

As the sample size increases, the upper bound on $p_\mathrm{int}$ asymptotically approaches the ground truth value of 0, as expected.  We show the 95\%-CI upper confidence bound as a function of the total charged time in Figure \ref{fig:totaltime_95ci} (solid line).  Diminishing returns are clearly visible.  While the upper limit can be efficiently constrained down to 12\% within 500 hours, a further improvement to 7\% requires a dramatic increase in observing time---from 483.7 to 2470 hours. This steep cost in the observing time is driven by both the statistics of the assumed binomial distribution, which would result in the occurrence rate constrained as $\sim 3/N$ à la ``rule of three'', and the fact that targets further down the ESM$_{15}$ ranking require more eclipses to achieve comparable constraints.

It is also worth noting that a optimized target selection is not strictly necessary to efficiently constrain the occurrence rate before diminishing returns set in.  The targets in this experiment were chosen based on rank ordering via ESM$_{15}$, which, if all planets are indeed bare rocks, provide the most efficient strategy to maximize the \textit{number} of observed targets.  Given the diminishing returns on occurrence rate and the higher ESM$_{15}$ targets being naturally favorable to observations, these results demonstrate that the strongest constraint realistically achievable will be naturally reached through a combination of the DDT and subsequent GO programs.

\subsection{Sensitivity to high albedo surfaces}

\begin{figure}[ht!] 
    \centering
    \includegraphics[width=0.96\columnwidth]{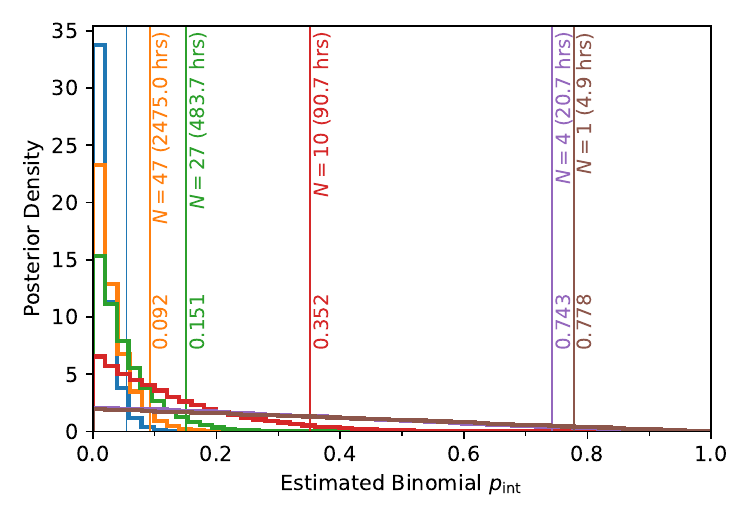}
    \caption{Same as Figure \ref{fig:posterior_probatmo}, but assuming that the planets are bare rocks with surfaces of $A_\mathrm{B}=0.2$ rather than blackbodies. The estimated 95\% CI are comparable to the blackbody case, and are robust unless the surfaces are brighter than $A_\mathrm{B} > 0.2$.}
    \label{fig:posterior_probatmo_bright}
\end{figure}

The results so far have assumed blackbody surfaces on all planets, such that a bare rock gives rise to the maximal eclipse depth.  A bright surface with a higher albedo can cause shallower eclipse depths \citep{mansfield19, coy24} and thereby reduce our ability to distinguish between a bare rock and an atmosphere.  As end member scenarios, we repeat the calculation now assuming that \textit{all} planets instead have a Bond albedo of 0.2 and 0.4, which roughly correspond to those of an ultramafic or an icy surface---the exact number depends how the stellar spectrum overlaps with the albedo spectrum of the surface.  We assume that the albedo does not depend on any specific axes.  We still use depths arising from blackbodies in calculating the observing time. 

In the intermediate Bond albedo of $A_\mathrm{B}=0.2$, the 95\%-CI upper limits are increased to 74.3\%, 35.2\%, 15.1\%, and 9.2\% 20.7, 90.7, 483.7, 2470 hours of observing time (Fig. \ref{fig:posterior_probatmo_bright}).  These limits corresponds to a roughly a 20\% increase compared to the blackbody case.  We stress that we are assuming \textit{all} targets have $A_\mathrm{B}=0.2$; this indicates that the upper limits are reasonably robust for moderate values of surface brightness. 

In the bright surface scenario of $A_\mathrm{B}=0.4$, the constraint is dramatically increased to 83\%, 61\%, 37\%, and 31\% for the four ESM cuts that correspond to 20, 100, 500, and 2500 hour surveys (Fig. \ref{fig:totaltime_95ci}).   We note that this end member scenario represents a somewhat extreme case that neither agrees with observations so far nor is predicted by theory \citep[e.g,][]{zieba23, coy24}.

\subsection{Robustness to prior assumptions}

\begin{figure*}[ht!] 
    \centering
    \includegraphics[width=0.8\textwidth]{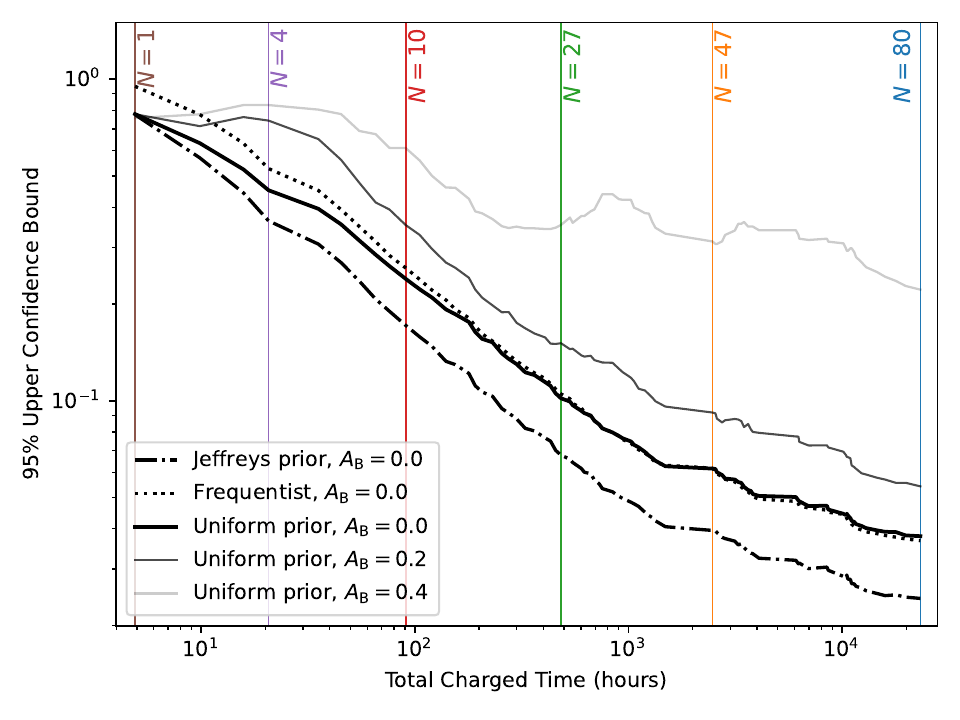}
    \caption{The obtained 95\% upper confidence interval (CI) as a function of the total (charged) observing time.  Here, the target list is generated by rank ordering planets by their ESM$_{15}$.  The baseline case, where we assumed a uniform prior and that all planets are bare blackbodies is shown as the solid line, which is also shown in Fig. \ref{fig:posterior_probatmo}.  The obtained CI using a Jeffreys prior (Beta(0.5,0.5)) and using (frequentist) Clopper-Pearson method is shown as dash-dot and dotted lines, respectively.  The obtained CI assuming all planets have brighter surfaces of Bond albedo 0.2 and 0.4 are plotted in fainter colors.  The vertical bars indicate the number of targets and the total number of hours based on different ESM$_{15}$ cuts, as in Fig. \ref{fig:posterior_probatmo}.}
    \label{fig:totaltime_95ci}
\end{figure*}

The obtained 95\% confidence upper bound for $p_\mathrm{int}$ compares well to that predicted by frequentist methods, but is predicated on an assumed prior.  We assumed a uniform prior on $p_\mathrm{int}$ for simplicity, but note that this is neither the objectively uninformative Jeffreys prior ($\beta(\frac{1}{2},\frac{1}{2})$ distribution) nor informed by intuition based on our understanding of planet formation.  Compared to a uniform prior, the Jeffreys prior assigns more weights to the values near 0 and 1, reflecting the fact that estimating probabilities near the extremes is harder.  To compute the posterior using the Jeffreys prior, we employ importance sampling and re-weight the samples derived from the uniform by the prior ratios.  We also compute the Clopper-Pearson interval (or ``exact'' interval) to calculate the confidence interval as would be estimated by a frequentist approach \citep{newcombe98_confidence}. 

We show the impact of the assumed prior in Fig. \ref{fig:totaltime_95ci}.  The uniform prior and the Clopper-Pearson (``Frequentist'') method agree as expected by construction, as a larger number of planets are observed.  There is less than 1\% difference with over 100 hours of observing time.  On the other hand, using the Jeffreys prior consistently finds the occurrence rate to be lower.  Specifically, with observing time of 500 hours, using the Jeffreys prior leads to an upper bound of 7\% rather than 12\% (green line).  This is expected analytically in the limit of 0 successes in a binomial distribution. 

We stress that, when estimating a parameter near an extreme, the general idea of ``\textit{priors do not matter for larger $N$}'' may not apply.  Even when the number of planets with detected atmospheres is not exactly zero, it is likely that the ratio of planets with detected atmospheres will still be close to zero, and the prior or the method of estimating $p_\mathrm{int}$ will still be relevant.   The Jeffreys prior case also serves as a Bayesian benchmark for how much one’s subjective belief about these planets having atmospheres can influence the conclusions drawn from observations.  We refer the readers to \citet{angerhausen25} for a comprehensive comparison of priors on occurrence rate constraints, wherein the authors performed a similar calculation as our experiment \# 1 but for occurrence of life.

\section{Results \#2: Target selection for the Cosmic Shoreline Hypothesis} \label{sec:result_cosho}

In this section, we present the results for how much evidence can be found for the Cosmic Shoreline hypothesis against the Random hypothesis, for the bolometric (\S \ref{subsec:testing}) and the XUV Cosmic Shorelines (\S \ref{subsec:testing_xuv}), using an optimized target list.  In the latter test, we also test how well the two Cosmic Shorelines can be distinguished from each other (\S \ref{subsubsec:testing_bol_xuv}).   Then, we repeat the experiment using the Pessimist hypothesis in place of the Random hypothesis (\S \ref{subsec:testing_pess}).  The summary of the results shown is provided in Table \ref{tab:summary_exp2}.

\begin{table*}[]
    \centering
    \begin{tabular}{c|c|c|c}
        \textbf{Injected} &  \textbf{Select targets to optimize for $\mathbb{E}[\Delta\mathrm{BIC}]$ between} & \textbf{Show histogram of $\Delta\mathrm{BIC}$ between} & Section \\ 
        \hline \hline
        \textbf{Bol. CS} & \textbf{Bol. CS} - \textbf{Random} & \textbf{Bol. CS} - \textbf{Random} & \S\ref{subsec:testing} \\
        \textbf{XUV CS} & \textbf{XUV CS} - \textbf{Random} & \textbf{XUV CS} - \textbf{Random} & \S\ref{subsec:testing_xuv}\\
        \textbf{XUV CS} & \textbf{XUV CS} - \textbf{Random} & \textbf{XUV CS} - \textbf{Bol. CS} & \S\ref{subsubsec:testing_bol_xuv}\\
        \textbf{Bol. CS} & \textbf{Bol. CS} - \textbf{Pessimist} & \textbf{Bol. CS} - \textbf{Random} & \S\ref{subsec:testing_pess} \\
        \textbf{XUV CS} & \textbf{XUV CS} - \textbf{Pessimist} & \textbf{XUV CS} - \textbf{Random} & \S\ref{subsec:testing_pess} \\
    \end{tabular}
    \caption{Summary of the results shown for \textbf{Experiment \#2} in \S \ref{sec:result_cosho}. Planets have atmospheres at random in the \textbf{Random} hypothesis, while no planet has an atmosphere in the \textbf{Pessimist} hypothesis. For each injected CS hypothesis, we use three values of $p_\mathrm{cs}=$0.33, 0.50, and 1.00, which is then an estimated parameter during recovery.}
    \label{tab:summary_exp2}
\end{table*}

\subsection{Testing for the Bolometric Cosmic Shoreline Hypothesis} \label{subsec:testing}

First, we present the results for how well the Cosmic Shoreline hypothesis can be distinguished from the Random hypothesis using the optimal set for 500 hours of observations.  We inject values of the occurrence rate of atmospheres on the wet side of the Cosmic Shoreline (with the only uncertainty from each target's uncertainty in $v_\mathrm{esc}$), with $p_\mathrm{cs}=1/3, 1/2$ and $1$. The case of $p_\mathrm{cs}$=1 has a plain division between dry and wet sides of the Cosmic Shoreline, while the boundary is less sharply defined for lower values of $p_\mathrm{cs}$. 

\begin{figure}[ht!] 
    \centering
    \includegraphics[width=0.99\columnwidth]{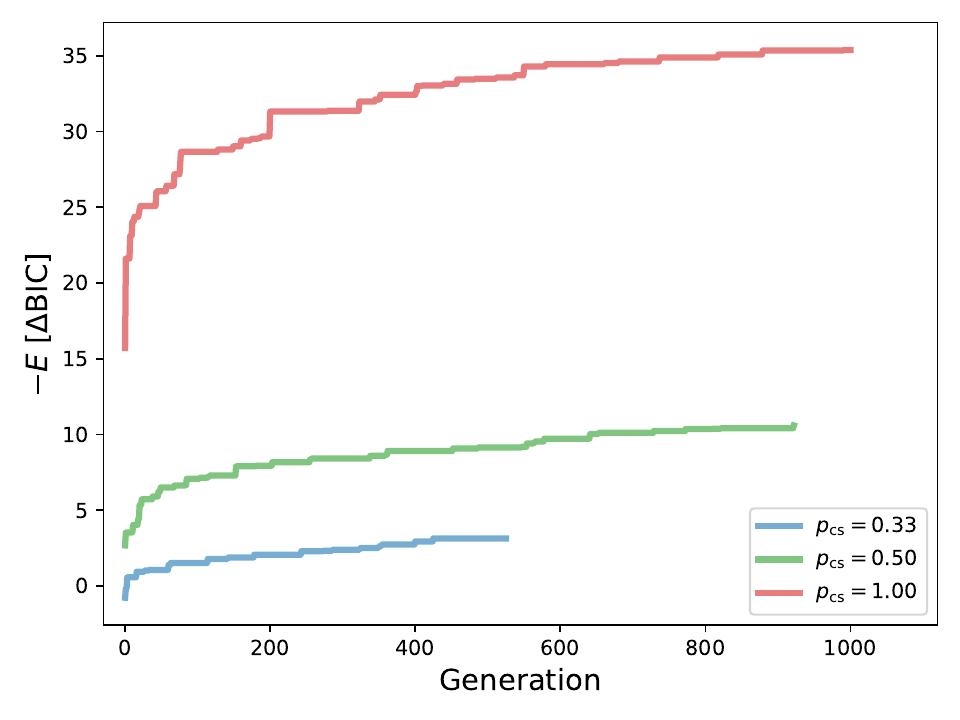}
    \caption{The fitness function, $-E[\Delta \mathrm{BIC}]$, plotted over generations for 3 different values of $p_\mathrm{cs}$.  Each generation is a population of 500 solutions, and the best fitness function in each generation is plotted.  Optimization is stopped when the fitness function does not improve over 200 generations.  The discontinuous jumps in the fitness function demonstrate the non-linear nature of the problem. }
    \label{fig:generation_converge}
\end{figure}

We show the convergence of the optimal solution in Figure \ref{fig:generation_converge}, in which the fitness function, the best value of $-\mathbb{E}[\Delta \mathrm{BIC}]$ in each population is plotted over generations.   Much of the convergence takes place earlier on as targets on the wet side of the Cosmic Shoreline are focused on, after which more gradual improvements occur iteratively.  

\begin{figure*}[ht!] 
    \centering
    \includegraphics[width=0.99\textwidth]{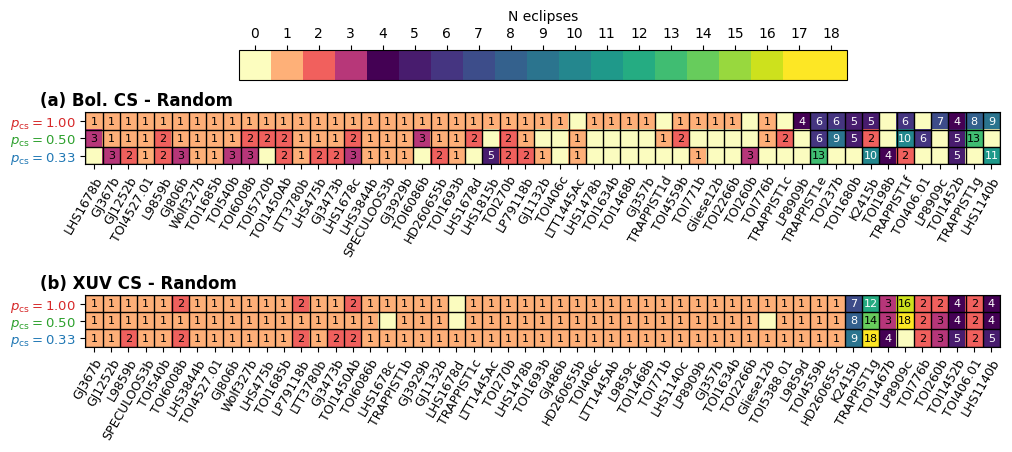}
    \caption{The optimal set of observations for the distinguishing \textbf{Cosmic Shoreline} versus \textbf{Random hypothesis} for 3 different values of $p_\mathrm{cs}=1, 1/2$ and $1/3$ for the \textbf{bolometric} (top panel) and the \textbf{XUV} Cosmic Shoreline hypotheses.  Planets have atmospheres at random with no particular trend in the Random hypothesis. The optimal set maximizes $-\mathbb{E}[\Delta \mathrm{BIC}]$ across 100 different Monte Carlo draws.  Targets that are not included at least once for the 3 values of $p_\mathrm{cs}$ are not shown.  In the top panel, there are 47, 38, and 32 total targets in each row; in the bottom panel, 52, 50, and 52.  Targets are ordered by the respective Priority Metric, such that targets further to the right are further on the wet side of the Cosmic Shoreline.  The injected Cosmic Shoreline passes through TRAPPIST-1 c for the bolometric and K2-415 b for the XUV.   }
    \label{fig:chromosome}
\end{figure*}

For all three values of $p_\mathrm{cs}$, the optimal strategy is a ``wide and shallow'' one: to establish a sufficiently wide baseline of planets on the dry side of the Cosmic Shoreline with one or two eclipses each--as they can be efficiently confirmed to be bare rocks--and stack eclipses on $\sim8$ targets on the wet side to test whether maximally hot bare rocks are ruled out (Figure \ref{fig:chromosome}).  

The large number of dry side targets is unsurprising, given that our formulation of the Cosmic Shoreline also predicts that the targets on the dry side will be bare rocks; therefore an observation of a bare rock on the dry side is as correct a guess as finding an atmosphere on the wet side.  We address an alternative strategy in \S \ref{subsec:testing_pess} and discuss the statistical approach in \S \ref{sec:discussion}. 

Importantly, the strategies for the targets on the wet side are generally consistent across different values of $p_\mathrm{cs}$. A modest variation between the three strategies is that, the more definitive the Cosmic Shoreline is, \textit{i.e. $p_\mathrm{cs}=1.00$}, the greater the number of targets and the fewer the number of eclipses per target.  In other words, as the chance of targets on the wet side being bare rocks increases ($p_\mathrm{cs} = 1/3$), it is more advantageous to bet and focus on a few high-yield systems.   This reflects the tradeoff between depth and breadth in facing uncertainty.

\begin{figure*}[ht!] 
    \centering
    \includegraphics[width=0.6\textwidth]{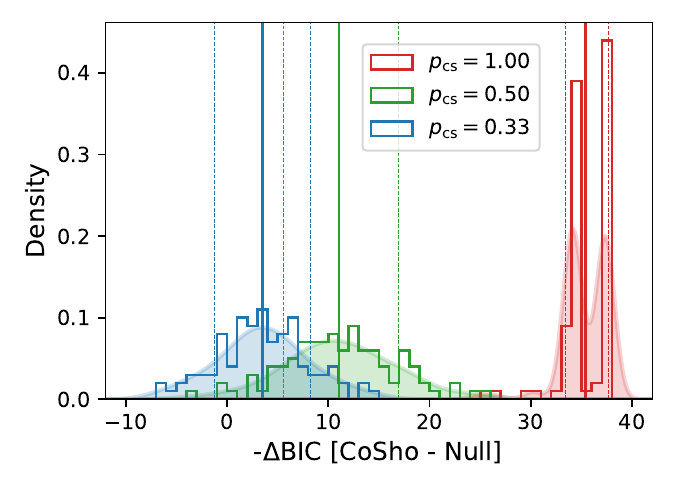}
    \caption{Histogram of $\Delta \mathrm{BIC}$ values between the \textbf{Cosmic Shoreline hypothesis} and the \textbf{Random hypothesis} across Monte Carlo draws for the $p_\mathrm{cs}=1, 1/2$ and $1/3$ case, in which planets on the wet side of the Shoreline have $p_\mathrm{cs}$ chance of having atmospheres.  The $\Delta \mathrm{BIC}$ values are obtained from the optimal strategy determined for \textit{each} $p_\mathrm{cs}$ realizations.}
    \label{fig:posterior_delta_bic}
\end{figure*}

Despite the stable strategy across three values of $p_\mathrm{cs}$, the resulting evidence for the Cosmic Shoreline is highly sensitive to the injected value of $p_\mathrm{cs}$.  In Figure \ref{fig:posterior_delta_bic}, we show the resulting histograms of $-\Delta \mathrm{BIC}$ values across 100 realizations (where each realization has varying draws of which targets have an atmosphere and what their atmospheric compositions are). 

Specifically, for the most definitive case $p_\mathrm{cs}=1.00$, plotted in red, the Cosmic Shoreline hypothesis is always favored with decisive evidence ($\Delta \mathrm{BIC}>10$) across all realizations, with resulting $-\Delta \mathrm{BIC} = 35.5^{+2.1}_{-2.0}$.  In other words, should the Solar system Cosmic Shoreline also apply to rocky planets around M stars unchanged, the hypothesis will always be favored after a 500 hour observing campaign.  

For the case of lower $p_\mathrm{cs}$, the resulting $\Delta\mathrm{BIC}$ is also more realization-dependent, resulting in a broad spread in the histogram. For the case of $p_\mathrm{cs}=1/2$, $\Delta \mathrm{BIC}$ shows both lower expected value and wider spread with $-\Delta \mathrm{BIC} = 11^{+6}_{-6}$.  In terms of confidence intervals, Cosmic Shoreline hypothesis will be favored with decisive evidence ($>$10) $\sim 58 \%$ of the realizations and with strong evidence ($>$5) $\sim 87 \%$.  

For the case of $p_\mathrm{cs}=1/3$, $\Delta \mathrm{BIC}$ shows a lower expected value still with $-\Delta \mathrm{BIC} = 4^{+5}_{-5}$, with confidence intervals of $\sim 8\%$ for decisive evidence ($>$10) and $\sim 37 \%$ of the realizations for strong evidence ($>$5).


To summarize, the optimal strategy is to observe a large number of both dry side and wet side targets, and is broadly stable across $p_\mathrm{cs}$ values. However, the resulting $\Delta$BIC values depend strongly on the injected occurrence rate of the atmosphere on the wet side, $p_\mathrm{cs}$ and the realization itself. Unsurprisingly, successfully distinguishing the Cosmic Shoreline hypothesis ultimately hinges on the true prevalence of atmospheres on the wet side of the Shoreline, even if an optimal set of targets were chosen.

\subsection{Testing for the XUV Cosmic Shoreline Hypothesis} \label{subsec:testing_xuv}

 We repeated the injection–recovery and GA optimization with the XUV Cosmic Shoreline Hypothesis. 
 
 The optimized observing strategy (Fig. \ref{fig:chromosome}) remains qualitatively unchanged: one to two eclipses on a broad dry‐side baseline and a concentrated stacking of four to eighteen eclipses on the wet side to test for potential atmospheres.  The number of eclipses on wet side targets increases as there are less wet side targets.  There is a considerable overlap between the target lists; 33 targets appear in both the bolometric target list (47 total) and the XUV target list (52 total).  Importantly, among the targets on the wet side in either of the list, there is an overlap of 9 targets total, with all but one target (TOI-1467 b) in the XUV wet side target list also appearing the bolometric list. 

\begin{figure}[ht!] 
    \centering
    \includegraphics[width=\columnwidth]{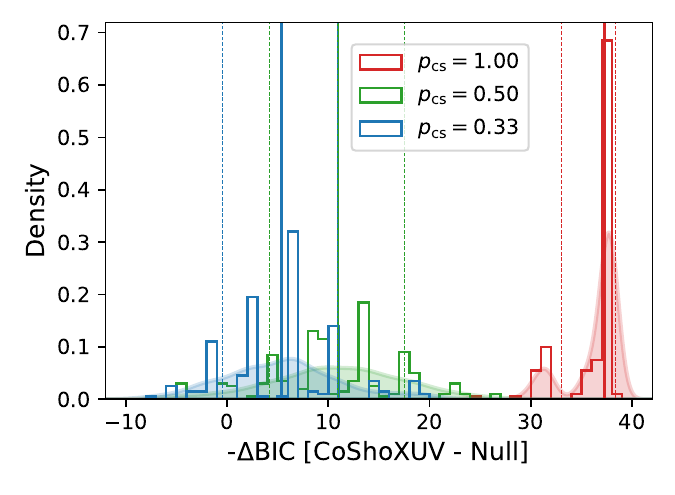}
    \caption{Same as Fig. \ref{fig:posterior_delta_bic} but between the \textbf{XUV Cosmic Shoreline hypothesis} and the \textbf{Random hypothesis}.  The $\Delta \mathrm{BIC}$ values are obtained from the optimal strategy determined for \textit{each} $p_\mathrm{cs}$ realization.}
    \label{fig:posterior_delta_bic_xuv}
\end{figure}

The resulting $\Delta$BIC distributions (Fig. \ref{fig:posterior_delta_bic_xuv}) closely track those found in \S \ref{sec:result_cosho}.  For an injected Shoreline with $p_{\rm cs}=1.0$, we recover a $-\Delta\mathrm{BIC}=37.2^{+1.2}_{-4.2}$; for $p_{\rm cs}=0.5$, $-\Delta\mathrm{BIC}=11^{+7}_{-7}$; and for $p_{\rm cs}=0.33$, $-\Delta\mathrm{BIC}=5^{+6}_{-6}$.  These values are within $\sim$10–15\% of their bolometric counterparts.  This demonstrates that the detectability of the M‐star Cosmic Shoreline is driven primarily by a target’s position relative to the empirical boundary rather than the choice of instellation metric. 

\begin{figure*}[ht!] 
    \centering
    \includegraphics[width=0.95\textwidth]{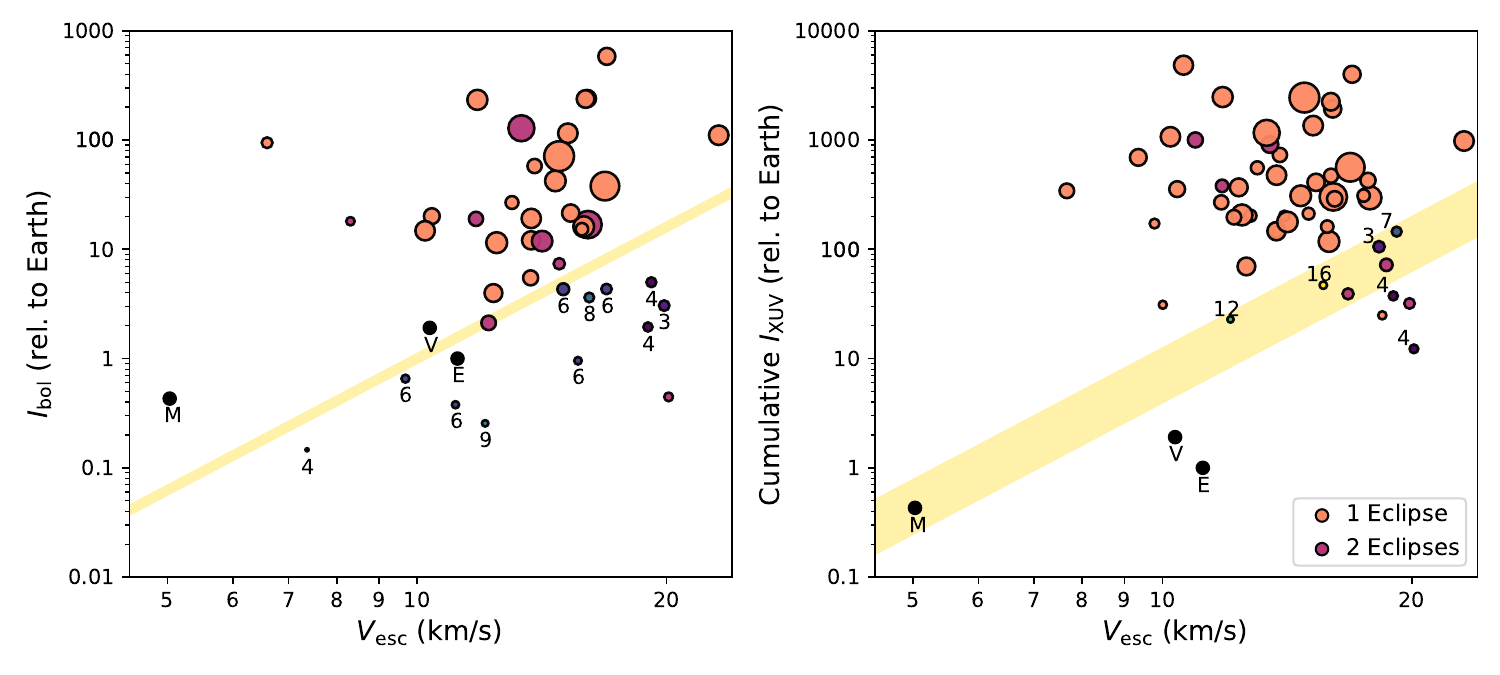}
    \caption{The optimal set of targets for maximizing the $\Delta$BIC between each \textbf{Cosmic Shoreline} and the \textbf{Random hypotheses}, using $p_\mathrm{cs}=1.00$ for the injected bolometric and XUV Cosmic Shorelines (top rows of Fig \ref{fig:chromosome} and Fig \ref{fig:chromosome}).   Targets not included are not plotted, and the number of eclipses are indicated only for targets with more than 2 eclipses.}
    \label{fig:cosho_100_both}
\end{figure*}

One notable difference is that the $\Delta$BIC values are more realization-dependent than the bolometric case, as seen by the modes in the histogram in Fig \ref{fig:posterior_delta_bic_xuv}.  This is likely due to the fact that there is a less reliable set of wet-side targets, as can be seen by the number of targets that are right on the upper edge of the Cosmic Shoreline in Figure \ref{fig:cosho_100_both}.  As such, whether these targets are on the wet side given the uncertainty in $v_\mathrm{esc}$ or not maintains a strong influence on the resulting $\Delta$BIC values.

Taken together, this suggests that bolometric instellation, in practice, provides a sufficiently good proxy for cumulative XUV exposure in population‐level inferences and that future work to refine XUV estimates, while necessary for other reasons (as discussed in \S \ref{sec:discussion}), will not substantially alter the core observational strategy outlined.

\subsubsection{Distinguishing the two Cosmic Shoreline hypotheses} \label{subsubsec:testing_bol_xuv}
\begin{figure}[ht!] 
    \centering
    \includegraphics[width=\columnwidth]{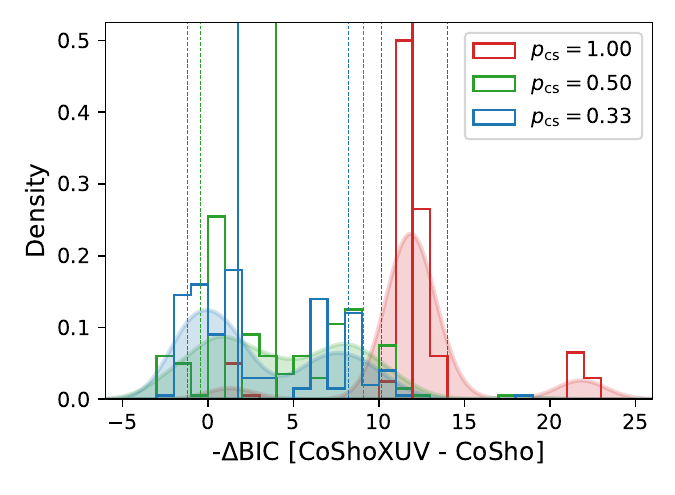}
    \caption{Same as Fig. \ref{fig:posterior_delta_bic} but between the \textbf{XUV Cosmic Shoreline }and the \textbf{bolometric Cosmic Shoreline hypothesis}, using the optimal target selection for $\mathbb{E}[\Delta\mathrm{BIC}]$ between XUV and the Random hypothesis.}
    \label{fig:posterior_delta_bic_xuvbol}
\end{figure}

Another question worth asking is whether we can find statistical preference between the two Cosmic Shorelines, in the case one of them was true. To quantify the preference between the two Shoreline definitions, we computed $\Delta\mathrm{BIC}{[\rm XUV-Bol]}$ using the optimized XUV target set for each value of $p_\mathrm{cs}$.   The resulting $\Delta\mathrm{BIC}$ is shown in Fig. \ref{fig:posterior_delta_bic_xuvbol}.   We find a median $\Delta\mathrm{BIC}=12.0^{+2.1}_{-1.8}$ for $p_\mathrm{cs}=$1.00, $\Delta\mathrm{BIC}=4^{+5}_{-4}$ for $p_\mathrm{cs}=$0.50,  and $\Delta\mathrm{BIC}=2^{+6}_{-3}$ for $p_\mathrm{cs}=$0.33.  Generally, except for the most definitive $p_\mathrm{cs}=1.00$ case of the XUV Cosmic Shoreline, only marginal evidence favoring XUV over bolometric Cosmic Shoreline is achieved, with a wide realization-dependent spread.

\subsection{Optimizing against the Pessimist hypothesis} \label{subsec:testing_pess}

\begin{figure*}[ht!] 
    \centering
    \includegraphics[width=0.75\textwidth]{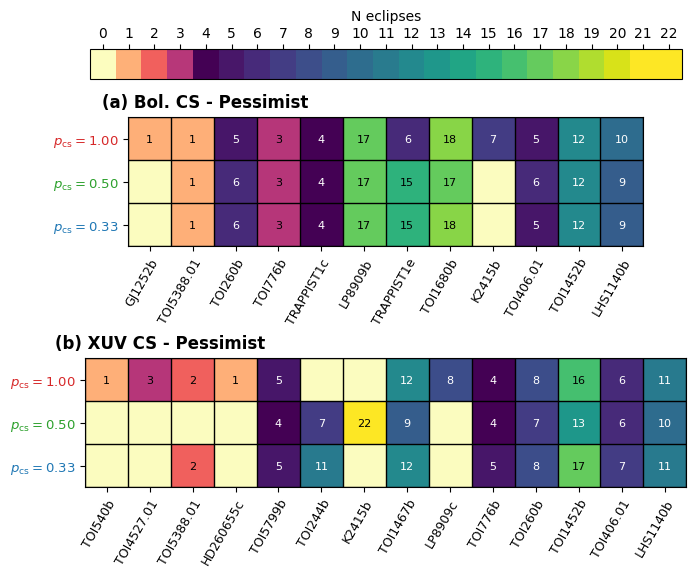}
    \caption{Same as Fig. \ref{fig:chromosome}, but for the \textbf{Cosmic Shoreline} hypotheses versus the \textbf{Pessimist hypothesis}, where in no planet has an atmosphere.  In the top panel, there are 12, 10, and 10 total targets in each row; in the bottom panel 12, 9, and 9.   Targets are ordered by the respective Priority Metric, such that targets further to the right are further on the wet side of the Cosmic Shoreline.  The injected Cosmic Shoreline passes through TRAPPIST-1 c for the bolometric and K2-415 b for the XUV.  Using the Pessimist hypothesis, rather than the Random hypothesis, as the null hypothesis for model to be compared to results in focusing more eclipses on the wet side of the Cosmic Shoreline. 
    }
    \label{fig:chromosme_pess}
\end{figure*}

\begin{figure*}[ht!] 
    \centering
    \includegraphics[width=0.95\textwidth]{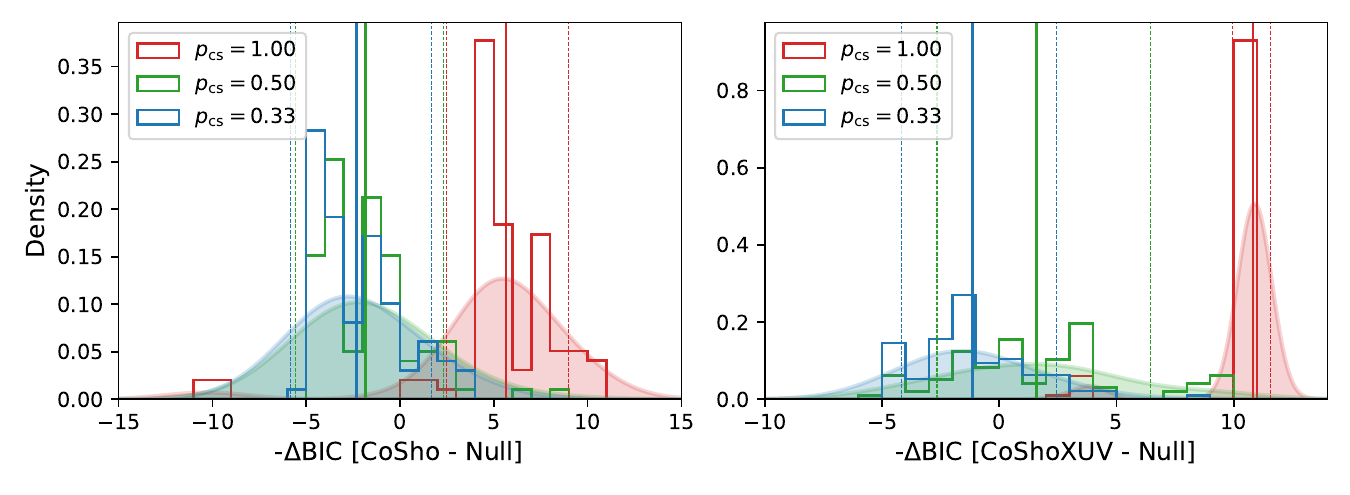}
    \caption{Same as Fig. \ref{fig:posterior_delta_bic} \& \ref{fig:posterior_delta_bic_xuv}, but using the optimized set that maximizes $-\mathbb{E}[\Delta\mathrm{BIC}]$ between the \textbf{Cosmic Shoreline} and the \textbf{Pessimist hypotheses}, plotted in Fig. \ref{fig:chromosme_pess}.  The $\Delta\mathrm{BIC}$ is still calculated against the \textbf{Random hypothesis}.}
    \label{fig:posterior_delta_bic_pess}
\end{figure*}

We repeat the optimization using the \textbf{Pessimist hypothesis} as the null hypothesis, rather than the Random hypothesis.  When optimizing against the Random hypothesis does indeed maximize the $-\mathbb{E}[\Delta\mathrm{BIC}]$, this ends up observing a large number of targets that are predicted to be bare rocks.  Since targets with atmospheres are of specific interest, an alternative is to use the Pessimist hypothesis as the null hypothesis against which the set of targets is optimized.  

We show the resulting target list in Fig. \ref{fig:chromosme_pess}.  For both the bolometric and XUV Cosmic Shoreline hypotheses, the optimized target list now consists of roughly one fifth as many targets, and instead stacks more eclipses on the wet side targets.  Importantly, there is still a good overlap between the optimized set for the bolometric and XUV Cosmic Shoreline hypotheses, with 7 of the targets on the wet side appearing in both lists.  

It is then of interest how well these target lists perform in terms of finding support for the Cosmic Shoreline.  We note that this must be done against the Random hypothesis rather than the Pessimist hypothesis that was used to optimize the target list.  The Pessimist hypothesis is trivially rejected in the frequentist sense; the Cosmic Shoreline is always favored with very strong support.  The Random hypothesis, in which planets have atmospheres with no particular trend, still provides a better baseline model to compare to if we wish to test whether a trend exists or not. 

We plot the resulting histograms of $\Delta \mathrm{BIC}$ against the Random hypothesis in \ref{fig:posterior_delta_bic_pess}.   For the bolometric Cosmic Shoreline, we achieve $-\Delta\mathrm{BIC}$ values of $6^{+3}_{-3}$, $-2^{+4}_{-4}$, and $-2^{+4}_{-4}$ for $p_\mathrm{cs}=$1.00, 0.50, and 0.33, respectively.  For the XUV Cosmic Shoreline, we achieve $10.8^{+0.9}_{-0.8}$, $2^{+4}_{-4}$, and $-1^{+3}_{-4}$ for $p_\mathrm{cs}=$1.00, 0.50, and 0.33, respectively. 

Similar to Fig. \ref{fig:posterior_delta_bic}, the resulting $\Delta \mathrm{BIC}$ is strongly realization dependent, and, unsurprisingly, finds worse $\Delta\mathrm{BIC}$ than when optimized specifically for the expected value.  In terms of how strong the support is,  for the bolometric Cosmic Shoreline, strong support ($\Delta\mathrm{BIC} > 5$) is found for more than 60\% of the realizations when $p_\mathrm{cs}=1.00$, and 5\% of the realizations when $p_mathrm{cs}=0.50$.   For the XUV Cosmic Shoreline, there is always a very strong support ($\Delta\mathrm{BIC} > 10$) when $p_\mathrm{cs}=1.00$, while there is a moderate tail (23\%) with strong support ($\Delta\mathrm{BIC} > 5$) when $p_\mathrm{cs}=0.50$.

Adding more observations of targets on the dry side (and if they are found to be bare rocks) will increase the $\Delta\mathrm{BIC}$ further.

\section{Discussion} \label{sec:discussion}

In this section, we identify the limitations of our statistical approach and discuss additional considerations for choosing targets.  We also address a number of issues related to population-level inferences using eclipse observations.

The most critical flaw of the statistical approach is that optimizing for targets this way may lead to conclusions that are \textit{merely} statistical.  That is, one might achieve the strongest possible population-level constraints while yielding \textit{only} weak individual results.

To illustrate this point, consider an extreme case: an observed sample that consists of 20 null results, each with exactly $q=prob(\mathrm{atmo})=0.5$, \textit{i.e.} effectively 20 coin tosses.  If one were to naïvely follow statistics, there is less than 1 in $10^6$ chance that all 20 targets are bare rocks.  This indicates a 4.8-$\sigma$ ``detection'' of at least one atmosphere in the sample; the Pessimist hypothesis is firmly rejected in the frequentist sense.  This is still so, even as we do not know which target has an atmosphere nor which one we should follow up on.  Such a conclusion, while statistically significant, obviously does not align with our intuition of what it means to detect something.  However, because our statistical framework does not quite capture this intuition---and as stacking eclipses necessarily yields diminishing returns---our methods, which emphasize statistical constraints, naturally push us toward these sort of unsatisfying conclusions borne out of a wide and shallow strategy.

One way to mitigate this tendency is by imposing a minimum threshold on detection significance.  However, setting an a priori detection threshold on potentially shallow eclipses ends up requiring an impractical number of eclipses per target---many of which may still turn out to be bare rocks.  As the opposite extreme of the previous example, consider a deep and narrow survey designed to achieve 4-$\sigma$ significances for discriminating 0.1 bar CO$_2$ atmospheres on the five best cumulative XUV Priority Metric targets (Figure \ref{fig:pm_esm}).  Such a survey needs, on average, 30-40 eclipses for each target and a total of over 2000 hours of charged time.  Clearly, a deep, narrow target selection approach that only prioritizes conclusively finding atmospheres on the wet side of the Shoreline does not sufficiently hedge the available observing time to produce a promising outcome either, as the sample size is too small to make a meaningful statistical statement and a large amount of time may be spent stacking eclipses on bare rocks.

\subsection{Additional considerations for target prioritization}

Given the caveat above, any statistical framework should inform but not dictate target selection; ultimately, target selection remains a hands-on process that demands per-target inspection and shrewd heuristics.  We discuss some of the considerations in our application that must be included this process.

\subsubsection{Host star characterization}

In the current work, the only host star property considered relating to atmospheric retention is the stellar effective temperature.  This is true for both the cumulative XUV scaling shown in Figure \ref{fig:cosho} and the early M star cut used in the injected hypothesis.  However, with observations, it will be beneficial (if not critical) to characterize the actual XUV environment of the planet in order to better understand atmospheric escape processes and the atmospheric photochemistry.  This is true of both the UV continuum and stellar flares.

If the planets are bare rocks, it is the high energy XUV irradiation that drives atmospheric loss.  While a snapshot of the current irradiation environment will not be a complete substitute for the total irradiation history, even a snapshot of the current XUV irradiation would allow for a more informed charting of the Cosmic Shoreline.  Moreover, flare rates of M stars and how they impact the long-term stability of an atmosphere remain open questions in the picture for atmospheric evolution and loss.

Instead, if any of the planets do have atmospheres, the probed atmospheres will have ongoing photochemical processes \citep{hu12_photochemistry}.  The UV flux from the host star is necessary to compute the photochemical steady-state of an atmosphere.  For secondary eclipse observations, this is generally less important for molecules, as the most prominent effect of photochemistry is to alter the abundance of O$_3$, which is not stable in warm atmospheres ($\gtrsim400$ K) and has no absorption features in the mid-infrared \citep{grenfell13, wunderlich21}.   Photochemical hazes may be of a greater concern, especially for the coldest planets, as they can have a strong forcing on the thermal structure of the atmosphere  \citep{peacock19, he20, ducrot23}.

Additionally, stellar flares can temporarily alter the composition of the atmosphere before the atmosphere equilibrates over the chemical timescale; and, depending on the flare rate, can have a more long-lasting effect on the atmospheric composition \citep{segura10, louca23}.  As such, knowing whether a flare has taken place during observation can help understand the atmospheric composition in its full context.

Not all host stars are amenable to UV characterization.  In practice, both observations of the UV continuum and characterizing flares can be challenging.  For most if not all of the M dwarf hosts being considered for a survey, there is no continuum observable in the FUV given the sensitivity of our instruments (HST/COS+STIS, which are the most sensitive in operation), with some continuum becoming observable in the NUV.  Additionally, MUSCLES survey suggest that M stars are not particularly predictable in the UV; \textit{i.e.} M stars with similar stellar properties ($R_\mathrm{s}$, $T_{\mathrm{eff}}$) can produce an order of magnitude difference in UV emission, making it difficult to scale the measured flux of an individual star to predict that of another \citep{france16, youngblood17}.

\subsubsection{Multiplanet systems}

The original Cosmic Shoreline in the Solar System delineates bodies with and without atmospheres that share a common irradiation environment as well as formation history.  In contrast, the M-star Cosmic Shoreline generalizes across diverse planetary systems with widely varying formation and evolution conditions.  Even if one could precisely determine the current XUV irradiation in each system, the M-star Cosmic Shoreline would still exhibit intrinsic scatter inherited from the stochastic nature of planet formation, shaped by heterogeneous conditions.


For this reason, observing targets within multiplanet systems offers the best opportunity to conduct a controlled experiment, in which the greatest uncertainty in the planets' birth environment --- the evolutionary history of the host star --- is shared among multiple planets.  This approach is particularly powerful if two or more planets in the same system straddle opposite sides of the fiducial Cosmic Shoreline.  Multiplanet systems with more than one rocky planet that are viable candidates include HD 260655, LTT 1445 A, and TRAPPIST-1.  Notably, some of these already have existing observations at 15 $\mu$m for at least one planet, providing a valuable foundation for comparative studies.  We note that, in principle, this consideration applies even if not all planets in the system are rocky.  After all, the Solar System’s Cosmic Shoreline encompasses both rocky and gaseous planets, as we discuss in further detail in \S \ref{subsec:already}.  This would further extend the set of relevant multiplanet systems to include \textit{e.g.} the TOI-406 and L98-59 systems. 


Moreover, observing targets in multiplanet systems also controls for one source of the uncertainty budget in interpreting observations.  Inferring that a planet is cold relative to the maximal temperature, \textit{i.e.} measuring $T_\mathrm{b}/T_\mathrm{max}$, relies on precise and accurate knowledge of the host star and system parameters. Determining $T_\mathrm{b}$ from a measured secondary eclipse depth requires knowing the spectrum of the star in the observed bandpass and the radius of the star, while computing $T_\mathrm{max}$ of a planet requires knowing the effective temperature of its host star (to complete the stellar spectrum) and the ratio of the semi-major axis of the planet’s orbit to the stellar radius ($a/R_\mathrm{s}$).  The statistical uncertainties in these parameters contribute a non-negligible portion of the uncertainty budget, comparable to the typical uncertainty in successful multi-eclipse observations \citep{xue24, mansfield24, Wachiraphan24}.  They are also susceptible to systematic errors that are difficult to quantify.  To first order, the uncertainties in the stellar spectrum, radius, and temperature are eliminated when comparing two sibling planets observed with the same instrument.  Therefore, planets in the same system that have different temperatures will allow the most unbiased and precise tests of recovering dayside temperatures observed, at leat in a relative sense.

\subsubsection{Unknown or imprecise mass and stale ephemerides} \label{subsec:mass}






In our methods, we made no explicit distinction between targets with directly measured masses and those with masses inferred from mass-radius relationships.  Masses of small exoplanets are typically measured via radial velocity (RV) measurements, but obtaining precise measurements for this sample is especially hard due to the intrinsically small amplitude and jitter induced by stellar activity \citep{ruh24_carmenes}.  When mass is not directly measured, one must rely on mass-radius relationships to estimate it from the observed radius, assuming an interior composition and hence bulk density --- or a prior distribution of plausible compositions.

The lack of a precisely known mass or a stale ephemeris introduces at least three challenges for a survey.  Firstly, if a target has a significantly lower density than a truly rocky planet, its interpretation in the context of atmospheric retention may be challenging, potentially invalidating its relevance to population-level hypothesis testing.  We note that this risk persists for directly measured mass (or radius) as well, given the precision usually achievable.  As a cautionary tale, for the planet TOI-1685 b, originally measured to have the density close to that of a water world, a combination of updated RV observations and JWST phase curve observations revealed that the planet was in fact ``rocky'' in composition and an airless body \citep{burt2024, luque24_dark}. 

Secondly, the lack of precise mass introduces an uncertainty in the position of a target relative to the Cosmic Shoreline along the $v_\mathrm{esc}$ axis.  For demonstration, if a planet has independently measured radius and mass with relative errors of 10\%, this leads to a propagated relative uncertainty in $v_\mathrm{esc}$ of 7\%.  However, if a mass-radius relationship must be invoked, assuming an uncertainty of $\sim$40\% in the prior of the unknown density leads to a propagated relative uncertainty in $v_\mathrm{esc}$ of 22\%.  This corresponds to roughly $\sim$40 targets with unknown masses that could in fact be on the other side of the assumed Cosmic Shoreline within 3 standard errors.

Finally, and perhaps most perniciously, the uncertainty in eccentricity and stale ephemeris contributes the most significantly to the uncertainty in the eclipse timing.  An eccentricity that is not precisely known therefore raises the possibility of missing the eclipse.  While most of the targets in our sample are expected to have nearly circular orbits ($e < 0.01$) due to tidal effects from their close-in orbits, even a slight eccentricity can introduce timing offsets, especially if a recent transit has not been observed to refine the ephemeris.  Given that the eclipse signal is often tangled with instrumental systematics, it is certainly possible that for small signal sizes that one cannot be certain that there is eclipse in the data at all \citep{august2024}.

Given the challenges posed by imprecise mass measurements, we emphasize the necessity of thoroughly characterizing eclipse observation targets through RV measurements beforehand, as well as frequent photometric observations of their transits to help maintain the precision of the ephemerides.   Targets whose masses and ephemerides are especially well-constrained should be prioritized, 


\subsubsection{MIRI systematics and repeatable eclipse depths} \label{subsec:systematics}



Early MIRI results have generally shown repeatable observations despite discernible time-dependent systematics in both LRS \citep[][e.g.]{mansfield24} and Imager \citep[][e.g]{greene23, zieba23, ducrot23, meier2025_toi1468b, fortune25}.  Repeatable eclipse depth measurements across multiple eclipses, as opposed to those from a single eclipse observation \citep[e.g.][]{xue24}, provide a rudimentary check that what we are measuring is a bona fide signal, rather than unmitigated systematics.  Perhaps as a cautionary tale, the two eclipses in \citet{august2024} did not repeat, with one eclipse indicating a shallow eclipse depth and the other a negative eclipse depth.  This was attributed to the presence of time-dependent systematics, clearly visible in the data. 

As the unobserved targets in the sample are of lower signal-to-noise than those observed in the first two cycles of JWST, repeating eclipses is a natural necessity in order to achieve the desired precision on the measured eclipse depths.  Here, the presence of time-dependent systematics may hinder scaling the errorbars as ideally as $\propto 1/{\sqrt{N}}$ as assumed in this study.

Despite such successes in measuring planetary thermal emission using time-series observations, it is important to note that the MIRI detectors were not designed for stable time-series observations at the level of precision required for exoplanet characterization.  As such, in order to correctly characterize secondary eclipses of small planets, there are time-dependent detector systematics to identify and remove. While still in the early years of JWST, we are rapidly gaining knowledge of both the origins and mitigation strategies for these signals.  The most prominent among these are a time-dependent exponential slope, which can be rising or decaying, seen at the beginning of time series observations \citep[e.g.][]{zhang24}, and which may have some dependence on the MIRI filter set in the filter wheel prior to observing \citep{fortune25}.  Practices for mitigating this initial ramp include removing the affected integrations or fitting the slope with an exponential function, both of which give promising results.  Another apparent source of time-dependent systematics is persistence from a cosmic ray striking one of the pixels in the stellar aperture  \citep[e.g. Holmberg et al. in prep][]{dicken24, fortune25_lhs1140c}.  Detecting and mitigating the effects of an unfortunately placed cosmic ray are possible if investigating the light curves at the pixel level.  A less tractable problem is the time-varying systematics throughout the observation that have as yet unidentified sources. There are a variety of such examples including broad time dependent features larger than the predicted eclipse depth \citep{august2024}, sudden drops in flux for no clear reason \citep{zhang24}, and short-period apparently periodic variations that can be fit with a Gaussian process but which have no clear origin (Allen+ submitted).

\subsection{Robustness of population-level inferences} 

Here we discuss the how robust the population-level inference can be against population-level false positives and negatives.

\subsubsection{Ubiquity of CO$_2$} \label{subsec:co2}

To start off, one assumption made in \S \ref{subsec:probatmo} is that we chose a 0.1 bar CO$_2$ model as a stand-in for all atmosphere models in determining the Bayes factor between an atmosphere and a bare rock.  Subsequent population-level inferences about $p_\mathrm{int}$ or $p_\mathrm{cs}$ now rest on the premise that 0.1 bar CO$_2$ atmosphere is a good proxy.  This is a defensible engineering choice, as it only takes an extremely small abundance of CO2 to create a significant spectral feature, and some CO$_2$ is a common feature in our simulated atmospheric compositions.

One caveat to this approximation still is that atmospheres with compositions or pressures that depart significantly from our fiducial model, such as a 1 bar O$_2$–N$_2$ atmosphere with minimal infrared opacity, could produce eclipse depths indistinguishable from a bare‐rock.  In such cases, one would undercount the number of true thick atmospheres, biasing $p_\mathrm{int}$ toward zero.

Nonetheless, we invoke the strength of the population-level approach here, namely in that we are more robust to outliers \citep{bean17_statistical}.   Such a transparent O$_2$-N$_2$ atmosphere with \textit{no} CO$_2$, while conceivable, would really pose a major concern only if we expect that M-star rocky planets were \textit{systematically} depleted in CO$_2$.   This is possible but unlikely, given the broad expectation from formation and evolution models, which instead predict its ubiquity \citep{hu20,kiteschaefer21, krissansen21}, nor from the Solar system planets, which all have CO$_2$ in amounts that would be detectable if they were in an exoplanet atmosphere.

\subsubsection{Population-level false positives} \label{pop_fp}

Continuing along this line of thought, it is worth pondering whether there could be \textit{systematic} effects that could lead to population-level false positives and negatives \citep{lustig19}.   For one, \citet{coy24} find a tentative 1D trend in brightness temperature vs.\ instellation, which could be explained with both geological processes such as space weathering or changes in surface grain size, as well as the onset of tenuous atmospheres (i.e., the Cosmic Shoreline).  Should the hypothesized trends continue to planets of even lower irradiation temperatures, such processes could potentially pose the threat of wrongly inferring the presence of a Cosmic Shoreline, even so as the targets in the trend were individually consistent with maximally hot bare rocks.  This is a separate problem from the robustness against \textit{generally }bright surfaces as explored in \S \ref{sec:result_occurence}. 

One potential mitigation strategy is to ensure that the target sample spans the full range of escape velocities, as to provide a leverage in inferring a true 2D trend.  The proposed geological processes that explain the 1D trend should generally not depend on the escape speed of the planet.  As such, searching for a trend with respect to escape speed \textit{at each given instellation}, could hint at the Cosmic Shoreline independent from geological processes.  We leave a statistical simulation using more realistic surface models for future work.  Additionally, we repeat for emphasis that any definitive individual detection of an atmosphere will have to utilize follow ups with phase curves or other instrument modes; in such cases, we expect breaking the degeneracy for individual targets with detailed observations would provide more insight into understanding geological processes.

\subsection{MIRI LRS vs MIRI F1500W}







As both MIRI LRS and MIRI photometry filters have been used to test for the presence of atmospheres on rocky planets, it is worth pondering whether a survey using MIRI LRS instead of MIRI photometry, or one that employs either or both (depending on the target) may be more advantageous.  For this purpose, the spectrum of LRS can be binned down to a single white light measurement, serving the same purpose as a broadband imaging filter to deduce the brightness temperature, as done in, e.g., \citet{mansfield24} and \citet{Wachiraphan24} \footnote[3]{Technically, this measurement should be called \textit{effective} temperature instead, given that LRS does not observed in a single narrow band and that it can capture most of the bolometric flux from the planet.}.  We discuss three salient points here: sensitivity of the two instrument modes, respective false positives and negatives, and whether there is any additional information to be gained from a \textit{spectral} characterization in the case of MIRI LRS.

Firstly, the choice of the 15-micron photometry filter of MIRI is optimized towards efficiently ruling out or detecting evidence of CO$_2$ absorption.  For this reason, while LRS used as a broadband instrument, in fact, achieves better signal-to-noise per eclipse in measuring the thermal flux of the planet for nearly all targets, MIRI 15-micron is still better at constraining whether there is a (CO$_2$-bearing) atmosphere or not.  This is unsurprising given that, (a) our definition of $prob(\mathrm{atmo})$ is based on how well a CO$_2$ atmosphere is ruled out; and (b) LRS--as a broadband instrument--probes both the continuum and absorption features, thereby the decreased eclipse depth due to the gas absorption is diluted by the continuum.  However, given that the ubiquity of CO$_2$ is a  robust prediction from planet formation point of view \citep{hu20}, a 15-micron photometric observation is still more efficient in terms of ruling out atmospheres, even if other gases (such as H$_2$O) are considered.

Secondly, LRS and F1500W measurements are affected slightly differently by potential sources of false positives and negatives, due to the difference in wavelength coverage.   One example of a false positive include bright surfaces \citep{hammond2024}, though this possibility is somewhat less than likely given the ubiquity of space weathering in the Solar system and the broad expectation that, if anything, space weathering will be more prominent on close-in rocky planets around M stars.  Here, we note that if LRS is used only as a white light instrument, it is not particularly more robust to bright surfaces, as their primary effect is the net cooling of the dayside, rather than imprinting any specific spectral features.  On the other hand, as an example of a potential false negative, thermal inversions caused by aerosols have been shown to adequately fit F1500W eclipse depth as well as a bare rock for TRAPPIST-1 b \citep{ducrot23}.  While this possibility requires somewhat a fine tuning of parameters, LRS could be more robust to this particular false negative due to the relative insensitivity to specific spectral features \citep{koll19, coy24}. 

Thirdly, LRS offers an additional spectral information that could be potentially used to characterize the composition of the atmosphere or even the surface mineralogy \citep{xue24, first2024, paragas25} .  For instance, for the planet GJ 1132 b, \citet{xue24} found that, while the broadband measurement is consistent with a Mars-like atmosphere within 1-$\sigma$, the emission spectrum and the computed $\chi^2$ based on the model spectra ruled out such atmosphere.  However, we stress that this form of spectral characterization can only really be used for targets with exceptional signal-to-noise.  As an illustrative comparison, for the planet LTT 1445 A b, the LRS spectra showed only a marginal improvement over the broadband in ruling out a thin CO$_2$-dominated atmosphere \citep{Wachiraphan24}.  Similarly, while there is intriguing possibility of spectrally characterizing the surface mineralogy with LRS, it is only really possible with numerous eclipses for the highest signal-to-noise such as LHS 3844 b \citep{paragas25}, and is not directly related to the central goal of finding out whether these planets as a population have atmospheres or not.

In searching for a population-level trend, whether it is of geophysical origin or due to atmospheres, having a consistent instrument choice is probably a good idea.  Given the broad expectation that most targets are likely to be bare rocks anyway, we suggest that the best strategy is to consistently use F1500W to rule out atmospheres.  Should any target be shown to be consistent with possessing an atmosphere, we can then utilize the full suite of observations (including LRS and phase curves) in GO follow-ups.

\subsection{Have we already found the M star Cosmic Shoreline?} \label{subsec:already}

\begin{figure*}[ht!] \centering
    \includegraphics[width=0.9\textwidth]{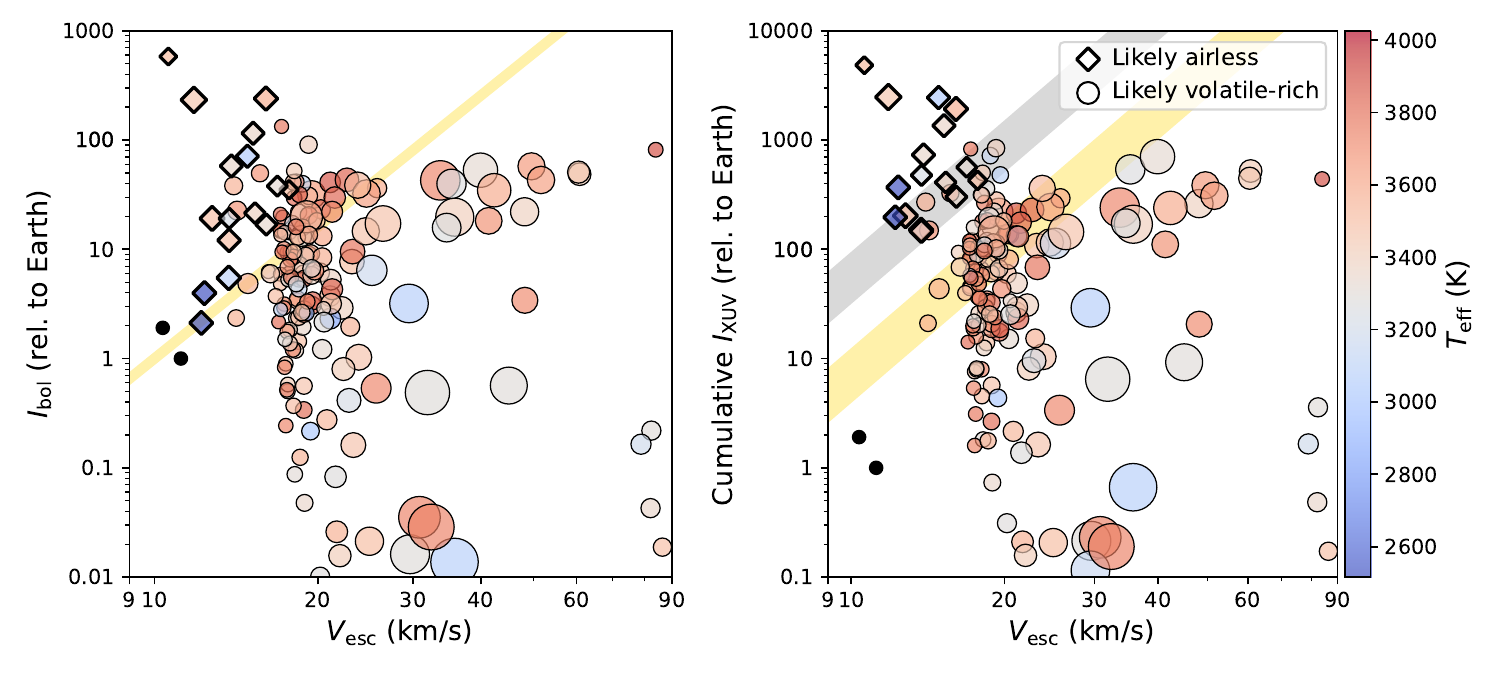}
    \caption{Cosmic Shoreline for planets---not limited to transiting rocky planets---around M stars for bolometric instellation (\textit{left panel}) and cumulative XUV instellation (\textit{right panel}).  All planets for which the presence of a thick ($\geq1$bar CO$_2$-bearing atmosphere has been ruled out from thermal emission observations are shown as diamonds, while all planets which are likely to be volatile-rich based on density measurements  ($\leq 0.6 \rho_\odot$) are shown as circles.  The markers are scaled by the inverse of their densities, with the size of the markers in the legend corresponding to Earth and Jupiter densities.  The yellow line in each panel is the same as Figure \ref{fig:cosho} and passes through TRAPPIST-1 c for the bolometric and above Mars for the cumulative XUV Cosmic Shorelines, respectively, and its width shows a representative uncertainty range in each instellation. The grey line is drawn by eye to suggest a loose separation within the observed planets, and its width represents the overlap between the two populations.}
    \label{fig:cosho_gas}
\end{figure*}

In the Solar System, the Cosmic Shoreline is not restricted to rocky planets but also applies to ice and gas giants, as well as small bodies such as moons and asteroids \citep{zahnle17}.  Although understanding habitability on rocky planets has been a key motivation for charting the Cosmic Shoreline around M stars, we can extend to other bodies as well insofar as we care about atmospheric escape mechanisms.  In Figure \ref{fig:cosho_gas}, we plot all planets around M stars for which an atmosphere has either been ruled out via emission measurements or inferred from density constraints, with a density cut applied to be less than 0.6 times that of the Earth \citep{Luque22}.  Here, planets relevant to habitability lie in the lower left corner, where there is a clear dearth of existing observations. 

Including non-rocky planets along the Cosmic Shoreline reveals a few noteworthy details.   Most importantly, there is no single line that \textit{definitively} separates bodies with and without atmospheres.  Regardless, in the escape speed-cumulative XUV instellation plane, there appears a somewhat messy separation by eye following $I \propto v_{\mathrm{esc}}^4$ between confirmed bare rocks and volatile-rich bodies.   We caution that this separation may well be artificial: the bare rocks towards the dry side are observationally favored due to higher temperatures, while our density cutoff selects low-mass and high-radius (and thus high-$v_\mathrm{esc}$) targets.  If anything, this potential bias makes the targets in the transitional regime particularly compelling targets for precisely constraining the Shoreline.   Secondly,  a handful of volatile-rich planets lie on the nominally dry side of the Cosmic Shoreline, implying either that the currently assumed Cosmic Shoreline (yellow line) is overly pessimistic or that sufficient diversity among M dwarf systems permits these planets to still retain their atmospheres.  In either case, this result offers some optimism regarding whether as-yet unobserved rocky planet targets would have atmospheres.  Finally, the Shoreline is much less noticeable in the escape speed-current bolometric instellation plane, indicating that focusing on the cumulative XUV as the main driver that carves out the Shoreline is indeed a sound choice \citep{berta25}.













\section{Summary and Conclusion} \label{sec:conclusion}

Answering the question \textit{Do rocky planets around M stars have atmospheres?} will be a lasting legacy of JWST.  To this end, the Cosmic Shoreline hypothesis, extrapolated from the Solar System, has provided a reasonable working hypothesis.  However, given the inhospitable M star environments for planetary atmospheres and that the M star Cosmic Shoreline is an amalgamation from diverse conditions for planet formation, we should not be so surprised if the metaphorical shoreline is in fact a desert with scattered puddles of stochastic origins.  The question of interest in this work is whether JWST observations can tell the difference.   
Towards this goal, we emphasize that \textit{precisely} stating this question as we do in \S \ref{sec:intro} is important in shaping expectations for what answers are possible, given limited resources, and for determining priorities in selecting targets. 

We have developed a fully reproducible, population-level framework for testing whether rocky exoplanets around M dwarfs have atmospheres and whether they follow an empirical ``Cosmic Shoreline'' separating airless bodies in instellation–escape speed space, for either bolometric of XUV instellation.  We took care in precisely defining the priors and the null hypothesis.

Then, by combining planet-formation outputs, 1D radiative-convective modeling, JWST/MIRI secondary-eclipse simulations, and genetic optimization, we simulate the population of rock planets around M stars and optimized target lists. From this, we demonstrate that:

\begin{itemize}
    \item \textbf{Occurrence‐rate constraints.} If all targets were dark bare rocks, with no dependence on escape speed or instellation, a survey of $\sim$27 rocky planets ($\sim$484 hours) would place a 95 \% upper limit of $\sim$12 \% on atmosphere occurrence, even without bespoke target selection.   Further, observing more targets only produces marginal gains ($\sim$7 \%) at a prohibitive time cost ($\sim$2470 h).  These constraints are robust against surfaces no brighter than $A_\mathrm{B}=0.2$.  This suggests that, over the lifetime of JWST, standard GO programs and the DDT will \textit{naturally} yield strong statistical upper bounds on whether or not all M-star rocky planets are bare rocks, \textit{if} they are indeed all bare rocks.

    \item \textbf{Evidence for the Cosmic Shoreline.} If an M-star Cosmic Shoreline akin to that in the Solar System exists, an optimized set of targets can find strong evidence for the trend ($-\Delta$ BIC $\geq$ 5) in a majority of realizations (87\%) for when the probability of a target on the wet side of the Cosmic Shoreline having an atmosphere, $p_\mathrm{cs}$, is $\gtrsim1/2$.  With low wet-side probabilities ($p_\mathrm{cs}\lesssim1/3$), however, the statistical distinction becomes strongly dependent on the realization, with $37\%$ of chance of a strong evidence ($\Delta$ BIC $\geq$ 5) for the Cosmic Shoreline.   

    \item \textbf{Observational strategy.}  The optimized strategy for obtaining the best statistical evidence for the Cosmic Shoreline is  a ``wide and shallow'' survey of $\sim50$ targets to statistically map out the instellation–escape speed plane comprising two to three eclipses on presumed ``dry'' planets combined with 4-18 eclipses on ``wet'' candidates.  This strategy maximizes the expected relative evidence between a trend in the prevalence of atmospheres and there being atmospheres at random.  However, this strategy also results in observing a large number of dry candidates; if this is to be avoided, a less wide approach of $\sim10$ targets focusing on wet side targets still finds moderate evidence for the Cosmic Shoreline.  From the results of either surveys, should any shallow eclipses hint at there being an atmosphere, detailed and focused follow-up via additional eclipses and phase-curve observations with broader wavelength coverage in GO cycles will robustly characterize individual atmospheres.

\end{itemize}






\begin{acknowledgements}

J.I. and E.M.R.K. acknowledge funding from the Alfred P. Sloan Foundation under grant G202114194.  J.I. was funded in part through support for JWST Program GO 3730, provided through a grant from the STScI under NASA contract NAS5-03127.  J.I. thanks the members of the DDT Scientific Advisory Council, Néstor Espinoza, Will Misener, Brandon Coy Park, and Natalie Allen for useful discussions.

\end{acknowledgements}

\bibliography{bib}{}
\bibliographystyle{aasjournal}

\end{document}